\pdfoutput=1
\documentclass{aastex}
\usepackage{spr-astr-addons}

\def\apj{ApJ\,}
\def\apjl{ApJ\,}
\def\aap{A\&A\,}
\def\mnras{MNRAS\,}
\def\pasj{PASJ\,}
\def\aj{AJ}
\def\apjs{ApJS}
\def\apss{Astrophysics and Space Science}
\def\pasp{PASP}
\def\POF{Physics of Fluids}
 
\RequirePackage{color}

\begin{document}

\title
{
The Physics of turbulent and dynamically 
unstable Herbig-Haro jets
}
\shorttitle{Herbig-Haro objects}
\shortauthors{Zaninetti}

\author{L. Zaninetti \altaffilmark{1}} 
\affil{
Dipartimento di Fisica Generale,   \\
Universit\`a degli Studi di Torino \\
Via Pietro Giuria 1,               \\
           I-10125 Torino, Italy }

\begin{abstract}
The overall properties of the Herbig-Haro objects 
such as centerline velocity ,
transversal profile of velocity ,
flow of mass and energy are explained
adopting two models for the turbulent
jet.
The complex shapes of the  Herbig-Haro objects ,
such as the arc in HH34 can be explained 
introducing the combination of different kinematic
effects such as velocity behavior along the main direction
of the jet and the velocity of the star in the interstellar
medium.
The behavior of the  intensity or brightness  of the line of emission
is explored in three different cases : transversal 1D cut ,
longitudinal 1D cut and 2D map.
An analytical explanation for the enhancement in  intensity or brightness 
such as usually modeled by the bow shock 
is given by a careful analysis of the geometrical 
properties of the torus.

\end{abstract}

\keywords{
 Pre-main sequence objects, young stellar objects (YSO's) and protostars (T
Tauri stars, Orion population, Herbig-Haro objects, Bok globules, bipolar
 outflows, cometary nebulae, etc.;
Jets, outflows, and bipolar flows 
}

\section{Introduction}

The Herbig-Haro objects , in the following HH ,
are  on the short distances from the central source
collimated patches of nebulosity associated 
with a central
very young star;
the name derives from the first two astronomer
which studied them in details , 
see \cite{Herbig1950,Haro1952}.
The HH are observed in the various band
of  the electromagnetic spectrum  :
in the radio , see 
\cite{Rodriguez1989,
Curiel1989,
Anglada1992,
Curiel1993,
Rodriguez1994} ;
in the infrared see \cite{Reipurth1997,
Reipurth1998,Chrysostomou2000,
Davis2000,
Davis2002,
Takami2005,
Takami2006};
in the optical see 
\cite{
Schwartz1988,
Rolph1990,
Scarrott1990,
Bohm1991,
Uchida1992,
Gomez1998,
Masciadri2001};
in the ultraviolet
\cite{Dopita1982,
Cameron1990,
Bohm1991,
Boehm1991,
Boehm1993} ;
in the X-ray  
see \cite{Pravdo1993,Raga2002}.
The HH's  are also  observed 
through emission-line spectra .
We remember that the astronomers observe both
the forbidden emission lines from low
ionization species , $[S II]$ 
and $[O I]$  as well as
ionization lines such as  $[O III]$  and $[Ne III]$, 
see  \cite{Hartigan2007}.

On long distances from the central source 
the HH are curved,
see 
 \cite{Salas1998,Bally2001,Bally2006}.
From a theoretical point of view the
apparent deceleration of HH34 has been modeled by
the 
interaction of a fragmented jet with 
the surrounding environment, see  
\cite{Cabrit2000}.
The theoretical problem of the curvature of the HH objects 
has been attached adopting 
 an HH jet/counterjet system that is immersed in an
isotropic stellar wind ,\cite{Raga2009}, or 
discussing the properties and similarities of
the laboratory and astrophysical flows,
see  \cite{Ciardi2008}
and  \cite{Hartigan2009}.
The already cited models concerning the HHs
leave a series of questions
unanswered or partially answered:
\begin {itemize}
\item Which are the laws of motion that regulate
 the propagation of HHs in the Interstellar Medium ~?
\item Is it possible to model the main physical
 properties of HHs such as matter 
 entertainment or mechanical luminosity~?
\item Can we model the bending of the HHs at the 
 light of the known rules of the kinematics~?
\item Can we model the  intensity or brightness  behavior along the 
 HHs using different models~?
\end{itemize}

In order  to answer these questions  Section~\ref{sec_hh}
reports the astronomical data of two HH objects.
Section~\ref{sec_tur} reports two physical theories 
on  turbulent jets which are converted in
astrophysical equations in Section~\ref{sec_tur}.
Complex phenomena such as jet's bending and train of knots
are explained in Section~\ref{sec_complex} adopting 
the composition of different kinematic effects
and the theory of the Kelvin-Helmholtz instabilities .
A set of models for the  intensity or brightness  of HH's which
arise from geometrical arguments are explored in 
Section~\ref{sec_image}.
Section~\ref{sec_image_turb} analyzes a model
for the  intensity or brightness  of HHs as given by 
a linear and a non linear relationship
between emission coefficient $j$  and turbulent power.

\section{The astronomical observations}

\label{sec_hh} 

This section presents the astronomical data 
of HH1 and HH34.
\subsection{The inner part of HH1}

The collimated jet HH1 has been observed in different
astronomical wavelengths such as 
$[FeII]$         by   \cite{Reipurth2000} ,
near infrared    by   \cite{Davis2000b} ,
far infrared     by   \cite{Molinari2002} ,
optical/infrared by   \cite{Nisini2005} ,
$H\alpha$        by   \cite{Riera2005} ,
$UV$             by   \cite{Li2007} and
$FeII$/$H2$      by   \cite{Podio2008}.

HH1 is part of the complex HH~1/2 that covers $3^{\prime}$ 
with its brightest components.
The distance of this complex is $460~pc$, see 
\cite{Molinari2002,Podio2008}.

The length of the jet , $L_{HH1}$ , measured from
the central source , VLA1 , to the knot A 
is  according to \cite{Nisini2005}
\begin{equation}
L_{HH1} = 22 ^{\prime\prime} =
0.049\;D_{460} \quad pc 
\quad ,
\label{lhh1}
\end{equation}
where $D_{460}$ is the distance in units 
of 460~$pc$ , see Table 2 in \cite{Nisini2005}.
The initial diameter $d_i$ (measured at the knot L-I)
and the final 
diameter $d_f$ (measured at the knot A )
are , according to Table 3 in
\cite{Nisini2005} 
\begin{eqnarray}
d_i = 2 \times 0.1 ^{\prime\prime} =0.000446 \;D_{460}\quad pc \nonumber \\
d_f = 2 \times 0.5 ^{\prime\prime} =0.00223\;D_{460} \quad pc
\quad .
\end{eqnarray}
The averaged radius of inner part 
HH1 is $\overline {r}=0.3 ^{\prime\prime}$ and 9 blobs 
characterize the structure.
The half opening angle , $\alpha/2$ , is 
\begin{equation}
\frac {\alpha}{2} =\arctan \frac{ d_f/2 - d_i/2} {L_{HH1}}=
0.0181~rad =1.041~^{\circ} 
\quad .
\end{equation}
With these data the laboratory parameter $x/d$ is
is
\begin{equation} 
\frac{x}{d} = \frac {L_{HH1}}{d_i} = 110 
\quad ,
\end{equation}
where $x$ represents the jet's 
length and
$d$ the nozzle's diameter~.

\subsection{HH34 the giant jet}

HH34 constitutes the archetypal bipolar collimated
jet from a young star and has been
carefully studied in 
deep $H\alpha$ and $[S II]$
with the Wide Field Planetary 
Camera 2 on board of the Hubble Space
Telescope and at the Gemini Observatory, see 
\cite{Reipurth2002,Beck2007}.

According to the data on HH34 as suggested
in \cite{Masciadri2002} and \cite{Reipurth1999}
the distance is $460~pc$,
the length of the jet (arc comprised)  , $L_{HH34}$ , is 
\begin{equation}
L_{HH34} = 1.5 \,pc \quad or \quad 4.62 \,10^{18}~cm 
\quad ,
\label{lhh34}
\end{equation}
the initial jet's diameter
is
\begin{equation}
d_i = 9.7 \,10^{-4} pc \quad or\quad 6 \times 10^{15}~cm 
\quad .
\end{equation}
The ratio $x/d$,
takes the value 
\begin{equation} 
\frac{x}{d} = \frac {L_{HH34}}{d_i} = 773 
\quad .
\end{equation}
A detailed study of the inner part of HH34 shows
a well collimated jet , see \cite{Beck2007} ,
which  means an opening angle of few degree ,
i.e. $\alpha=2.86 ^{\circ}$ (measure of the author).
Making reference to Figure~3 in
\cite{Reipurth2002}
the inner part of HH34 has  
an  averaged radius of $0.5 ^{\prime\prime}$,
a   length of  $27^{\prime\prime}$ 
and
twelve blobs . 

\section{The turbulent jet}
\label{sec_tur}

This Section reports known and new formulas
on turbulent jets.
We selected the modern approach to
turbulent round jets, see  
\cite{Pope2000} and the classical approach 
of \cite{landau}. 
These two approaches were already used to model
the extragalactic jets , see   
\cite{Zaninetti2007_b,Zaninetti2009d}.

\subsection{The exact solution}
\label{turbulent}

The theory of turbulent round jets can be 
found in different textbooks.
The more important formulas are now reviewed 
as extracted 
from chapter~V in  \cite{Pope2000} ;
similar results can be found in \cite{foot}
and in \cite{Schlichting2004}.
We start with the centerline velocity $U_0(x)$ 
, equation~(5.6) in \cite{Pope2000} , 
as measured 
in the laboratory experiments :
\begin{equation}
\frac{U_0(x)} {U_1} = \frac {B}{x/d} 
\quad ,
\label{vlab}
\end{equation}
here $x$ denotes the main direction , 
$d$ is the diameter of the nozzle,
$B$ is a constant derived in the laboratory
that takes the value 5.8, and
$U_1$ is the initial jet's velocity.
The solution of the mean velocity $<U>$ 
, equation~(5.100) in \cite{Pope2000} ,
along the main direction is
\begin{equation}
<U> =U_1 8 a_T \nu_T \bigl (\frac {d}{x} \bigr)  \frac{1}{ (1 +a_T \eta^2)^2}
\label{uturb} 
\quad ,
\end{equation}
where $\eta=\frac{r}{x}$ , 
$r$ is the radius of the jet at $x$,
$a_T$ is a constant and 
$ \nu_T$ is the turbulent viscosity.
The viscosity
,
equation~(5.104) in \cite{Pope2000}
,
 is
\begin{equation}
\nu_T = \frac{S}{8(\sqrt{2} -1)} 
\quad ,
\end{equation}
and $a_T$, equation~(5.18) in \cite{Pope2000},
 is
\begin{equation}
a_T = \frac{(\sqrt{2} -1)} {S^2}
\quad ,
\end{equation}
where $S$ 
is connected with the opening angle 
$\alpha$ through the following relationship 
\begin{equation}
S = \tan \frac {\alpha}{2}
\quad .
\end{equation}

The production of turbulent kinetic energy 
in the boundary layer approximation 
,
equation~(5.145) in \cite{Pope2000} ,
is
\begin{equation}
{\mathcal{P}} = {\nu_T (\frac{\partial <U>}{\partial y }})^2
\quad ,
\label{pturb}
\end{equation}
where $y$ is a Cartesian coordinate 
that can be identified 
with $r$, the perpendicular distance from the centerline
and the  units are $\frac {erg}{s\, cm^3}$ .
The flow rate of mass $m(x)$ is , see equation~(5.68) 
 in \cite{Pope2000} ,
\begin{equation}
\dot {m}(x) = 2 \pi \rho ( b_{1/2} (x))^2 U_0(x) 
\times \int_{0}^{\infty} \xi f(\xi) d \xi
\quad ,
\label{flowmass}
\end{equation}
where 
\begin{equation} 
\xi = \frac{r} {b_{\frac{1}{2}}(x)}
\quad ,
\end{equation}
and 
\begin{equation}
f (\xi) = \frac{1}
{ 
(1 + A \xi^2)^2
} 
\quad ,
\end{equation}
where $A$ is a constant that will be later defined
and $b_{\frac{1}{2}} $ is the value of the radius at which 
the velocity is half of the centerline value.
The jet draws matter from the surrounding mass of fluid.
Hence, the mass of fluid carried by the jet increases 
with the distance from the source.
The flow rate of kinetic energy $E(x)$ is , see equation~(5.69) 
 in \cite{Pope2000} ,
\begin{equation}
\dot {E } (x) = \frac{ \pi \rho}{ b_{\frac{1}{2}}(x)} (b_{\frac{1}{2}}(x) U_0(x))^3 
\times \int_{0}^{\infty} \xi (f(\xi))^3 d \xi
\quad ,
\label{flowenergy}
\end{equation}
which has units of $\frac {erg}{s\, cm^3}$ .
 
\begin{figure}
 \begin{center}
\includegraphics[width=7cm]{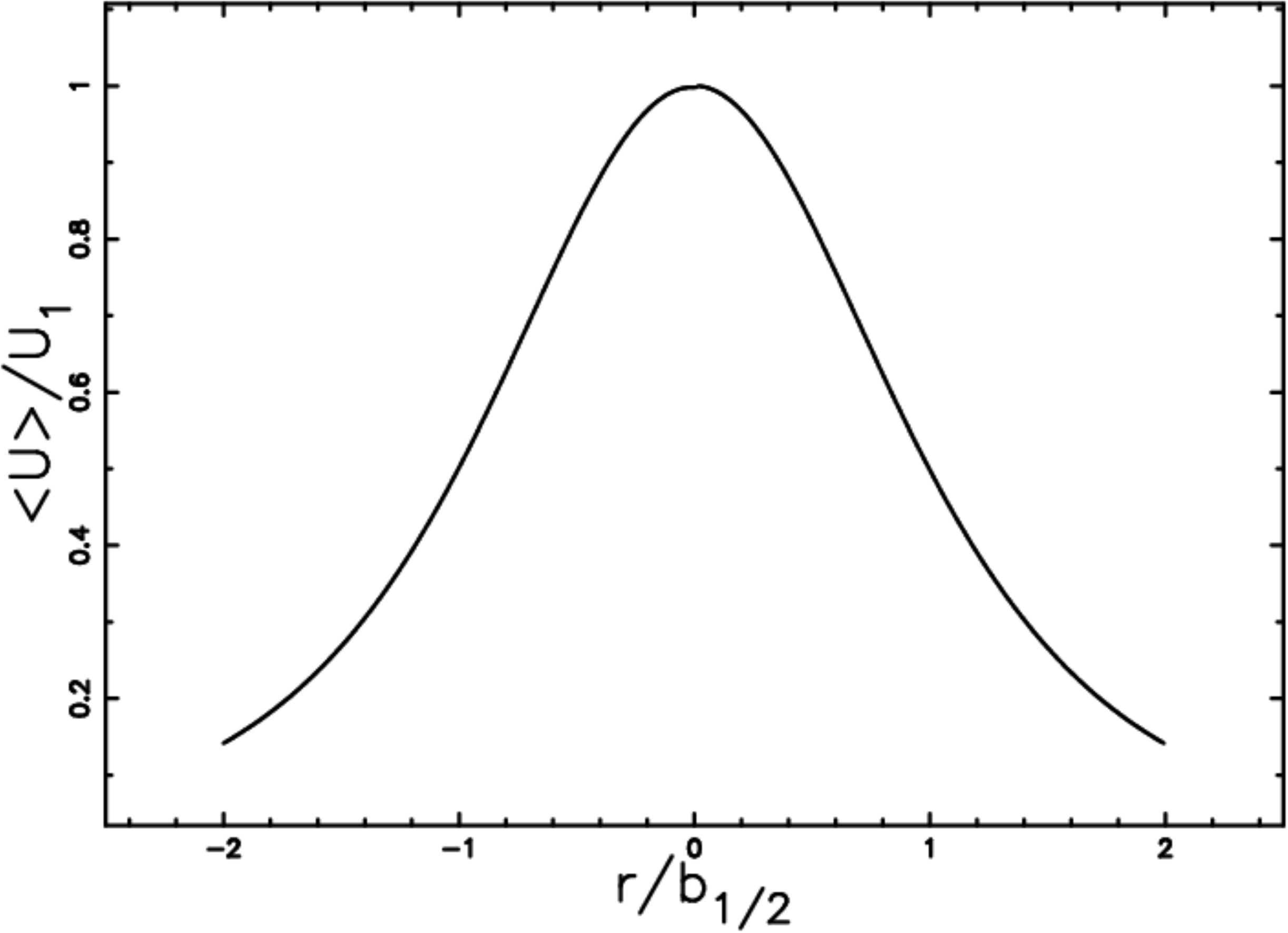}
 \end {center}
\caption {
Mean velocity profile vs. channel radius 
in a turbulent round jet.
The velocity distribution is a function growing from zero 
(at the wall of the channel) 
to a maximum value in the central region.
The experimental data at $Re$ $\approx$ $10^5$ 
and $x/d$= 40 , 60, 75 and 97.5 can
be found in Wygnanski~Fiedler~1969.
 }%
 \label{prof_velocity}
 \end{figure}

The previous formulas are exactly the same as 
in \cite{Pope2000};
we now continue toward the astrophysical applications.
The self-similar solution for the velocity , 
equation~(\ref{uturb}) , can be re-expressed
introducing the half width $x=b_{\frac{1}{2}} /S $ 
\begin{equation}
<U> =U_1 8 a_T \nu_T \bigl (\frac{d} {x} \bigr)  \frac{1}{ (1 +A(\frac{r}{b_{\frac{1}{2}}})^2)^2} 
\quad ,
\end{equation}
where $A=\sqrt{2} -1 $~.
From the previous formula is clear 
the universal scaling of
the profile in velocity 
that is reported in Figure~\ref{prof_velocity}.

From a careful inspection of the previous formula it is clear
that the variable $x$ should be expressed in $d$ units 
(the nozzle's diameter) in order to reproduce 
the laboratory results. In doing so we should 
find the constant $k$ that allows us to deduce $B$ 
\begin{equation}
B = k \times 8 a_T \nu_T 
\quad .
\end{equation}

Table~\ref{data} reports a set of $S$ , $B$ and $\nu_T$ 
for different opening angles $\alpha$ .
\begin{table}
 \caption {Parameters of the turbulent jet
 when $k$ = 0.54 . }
 \label{data}
 \[
 \begin{array}{ccccc}
 \hline
 \hline
 \noalign{\smallskip}
\alpha [rad] & \alpha [degree] & S & B & \nu_T \\
 \noalign{\smallskip}
 \hline
 \noalign{\smallskip}
0.035 & 2 & 0.017 & 30.96 & 0.0052 \\
0.087 & 5 & 0.043 & 12.37 & 0.013 \\
0.14 & 8 & 0.07 & 7.72 & 0.021 \\
0.185 & 10.64 & 0.093 & 5.79 & 0.028 \\
0.261 & 15 & 0.131 & 4.1 & 0.039 \\
0.343 & 20 & 0.17 & 3.06 & 0.053 \\
0.436 & 25 & 0.22 & 2.43 & 0.067 \\
0.523 & 30 & 0.26 & 2.01 & 0.08 \\
\noalign{\smallskip}
\noalign{\smallskip}
 \hline
 \hline
 \end{array}
 \]
 \end {table}
The assumption here used is that $k$ is the same for
different angles.
The velocity expressed in these practical units 
is 
\begin{equation}
<U> =B U_1 \frac{d}{x} 
\frac{1}
{ 
(1 +A~(\frac{r}{b_{\frac{1}{2}}})^2)^2
} 
\label{uturbpract}
\quad .
\end{equation}

This formula can be used for $x$ expressed in $d$-units 
when $x>B$.

The first derivative 
of the profile in velocity as given by formula~(\ref{uturbpract})
 with respect to the radius 
 is 
\begin{eqnarray} 
\frac {d}{dr} <U> = \nonumber \\
 U_1 \times
\frac
{
-4\, \left( \sqrt {2}-1 \right) {\it A~ k}\,{{\it {b_{\frac{1}{2}}}}}^{6}A~r
d}
{
\tan \left( \frac{\alpha}{2} \right) \left( \sqrt {2}-1 \right) x \left( {
{\it {b_{\frac{1}{2}}}}}^{2}+A~{r}^{2} \right) ^{3}{{\it {b_{\frac{1}{2}}}}}^{2}
}
\quad .
\end{eqnarray}

The production of turbulent kinetic energy is 
\begin{eqnarray}
{\mathcal{P}} =
 \nu_T U_1^2 \times 
\nonumber \\
\times \frac
{
2\, \left( \sqrt {2}-1 \right) ^{2}{{\it A~ k}}^{2}{{\it {b_{\frac{1}{2}}}}}^{12}{A~
}^{2}{r}^{2}  d^2
}
{
\tan \left( \frac{\alpha}{2} \right) \left( \sqrt {2}-1 \right) ^{3}{x}^{2
} \left( {{\it {b_{\frac{1}{2}}}}}^{2}+A~{r}^{2} \right) ^{6}{{\it {b_{\frac{1}{2}}}}}^{4}
}
\quad .
\label{powerturb}
\end{eqnarray}
It is interesting to note that the maximum of 
$ \mathcal{P} $ , is at 
\begin{equation}
r= \frac{1}{\sqrt{5}} \frac{b_{\frac{1}{2}}} {\sqrt{A~}}
=0.69 {b_{\frac{1}{2}}}
\quad . 
\end{equation}

In these practical units the flow rate of mass ,
equation~(\ref{flowmass}), becomes 
\begin{equation}
\dot {m }(x) =
\frac{
\pi \,\rho\,x \left( \tan \left( \frac{\alpha}{2} \right) \right) ^{3}{
\it U_1}\,d
}
{
\sqrt {2}-1
}
\quad ,
\label{mass_practical}
\end{equation}
and flow of kinetic energy , equation~(\ref{flowenergy}) ,
is 
\begin{equation}
\dot {E}(x) = 
\frac
{
{d}^{3}{{\it U_1}}^{3} \left( \tan \left( \frac{\alpha}{2} \right) \right) 
^{5}\rho\,\pi 
}
{
10\, \left( \sqrt {2}-1 \right) x
}
\quad .
\label{energy_practical}
\end{equation}
A more sophisticated approach 
makes extensive use of 
a statistical
mass-averaging technique for compressible turbulent flow,
see \cite{Favre1969,Bicknell1984}.

\subsection{The simple solution}

We now outline the conservation of the momentum flux
in a "turbulent jet"  , see \cite{landau} .
The initial point is characterized by the following
section 
\begin {equation}
A_0=\pi~r_0^2
\quad .
\end{equation}
On introducing $\alpha$ ,the opening angle ,
$x_0$ ,the initial position on the $x$--axis, and
$v_0$ , the initial velocity ,
the section $A$ at position $x$ is
\begin {equation}
A(x)=\pi \bigl ({\it r_0}+ \left( x-{\it x_0} \right) \tan \left(\frac{\alpha}{2}
 \right) \bigr )^2
\quad .
\end{equation}
The conservation of the total momentum flux states that
\begin{equation}
\rho v_0^2 A_0 =
\rho v(x)^2 A(x)
\quad ,
\label{conservazione_simple}
\end {equation}
where $v(x)$ is the velocity at position $x$ .
Due to the turbulent transfer, the density $\rho$
is the same on both the two sides of
equation~(\ref{conservazione_simple}).
The trajectory of the jet as a function of the time
is easily deduced
from equation~(\ref{conservazione_simple})
\begin {eqnarray}
x= 
\nonumber \\
{\frac {{\it x_0}\,\tan \left(\frac{\alpha}{2} \right) - {\it r_0}+
\sqrt {{\it r_0}\, \left( {\it r_0}+2\,\tan \left(\frac{\alpha}{2}
 \right) {\it v_0}\,t \right) }}{\tan \left(\frac{\alpha}{2} \right)
}} \quad .
\label{traiettoria_simple}
\end{eqnarray}

The velocity as function of the time  turns out to be
\begin{equation}
{\it v(t)}={\frac {{\it v_0}\,{\it r_0}}{\sqrt {{\it r_0}\, \left( {\it r_0}
+2\,\tan \left(\frac{\alpha}{2} \right) {\it v_0}\,t \right) }}}
\quad .
\label{velocita_simple}
\end {equation}

The flow rate of mass 
and 
kinetic energy
are respectively 
\begin{equation}
\dot {m}(x) = \pi r^2 \rho v 
\quad ,
\label{flowmass_simple}
\end{equation}
\begin{equation}
\dot {E}(x) = \pi r^2 \frac{1}{2} \rho v^2 
\quad ,
\label{flowenergy_simple}
\end{equation}
where $r$ and $v$ are the momentary 
radius and velocity of the jet.

\section{The physics of HH's}

\label{sec_phys}
This Section reports the centerline velocity,
the equation of motion ,
the flow of mass and the
flow of energy for the two turbulent models 
here considered.

\subsection{The exact solution}

Equation~(\ref{vlab}) allows us to 
deduce the centerline velocity of the 
turbulent astrophysical jet
\begin{equation}
u_0(x) = 
\frac
{
0.5409\,{\it u_{100}}\,{\it d_1}
}
{
\tan \left( 0.00872\,{\it \alpha_{deg}} \right) {\it x_1}
}
\frac{Km}{sec}
\quad , 
\label{u_astro}
\end{equation}
where $\alpha_{deg}$ is the opening angle expressed in degree,
$u_{100}$ is the initial velocity expressed in units 
of $100\frac{km}{s}$ ,
 $ u_{100}= \frac{U_1 [km/s]}{100} $
,
$d_1$ is the diameter of the nozzle in $pc$ units 
and 
$x_1$ is the length of the jet in $pc$ units. 

\begin{figure}
 \begin{center}
\includegraphics[width=7cm]{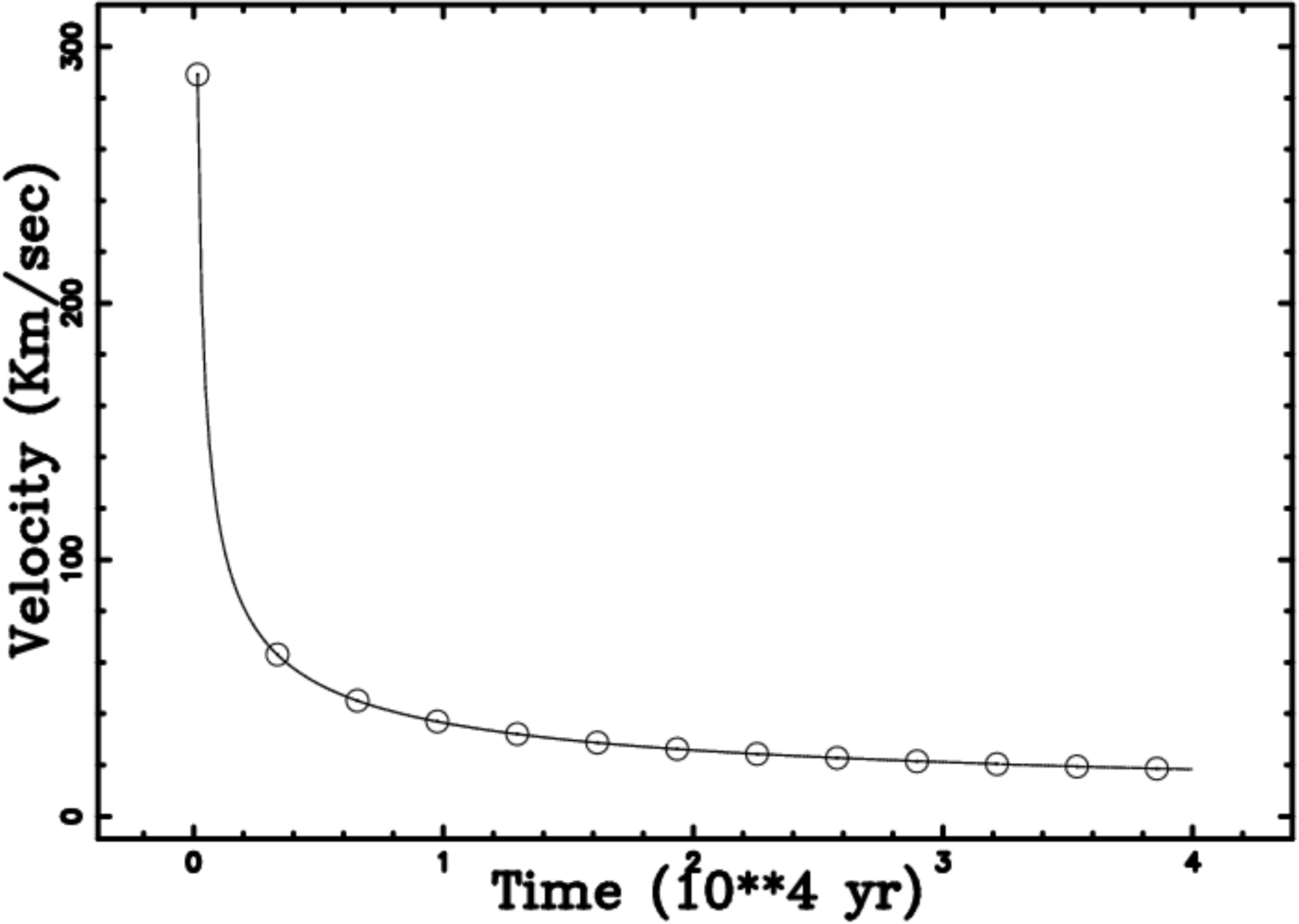}
 \end {center}
\caption {
Velocity of HH34 in $\frac{Km}{s}$ 
versus time in $t_4$ units when
$\alpha =2.86 ^{\circ} $ ,
$u_{100} =6.5$ ,
$d_1 = 1.94 \,10^{-3}$ 
and 
$x_1 = 1.49 $.
 }%
 \label{velocity_hh34}
 \end{figure}

The previous equation allows us to deduce 
the equation of motion for a turbulent astrophysical jet ,
\begin{equation}
x(t) =
1.050\,\sqrt {{\frac {{\it u_{100}}\,
{\it d_1}\,{\it t_4}}{\tan \left( 
 0.00872\,{\it \alpha_{deg} } \right) }}}
\; pc
\quad ,
\label{x_astro}
\end{equation}
where $t_4$ = $t[yr]/(10^4)$ .
The radius of the turbulent jet is 
\begin{equation}
r(x) = [\frac {d_1}{2} + x(t) \tan (\frac {\alpha_{deg}} {2} ) ] pc
\quad .
\label{r_astro}
\end{equation}
Combining equations~(\ref{u_astro}) , 
 (\ref{x_astro}) 
and (\ref{x_astro}) is 
possible to deduce the velocity of the HH object ,
for example HH34 , as function of the time 
see Figure~\ref{velocity_hh34}.

The power released in the turbulent cascade is
\begin{eqnarray}
\epsilon(r,x) = 
\nonumber \\
\frac {
2\, \left( \sqrt {2}-1 \right) ^{2}{{\it k}}^{2}{{\it b_{\frac{1}{2}}}}^{12}{A~
}^{2}{r}^{2}
}
{
\tan \left( \frac{\alpha}{2} \right) \left( \sqrt {2}-1 \right) ^{3} \left( {{\it b_{\frac{1}{2}}}}^{2}+A~{r}^{2} \right) ^{6}{{\it b_{\frac{1}{2}}}}^{4}
}
\nonumber \\
\times 
(\frac {d_1}{x_1})^2 
\label{e_astro}
\quad .
\end{eqnarray}
The flow rate of mass , see equation~(\ref{mass_practical}) ,
as expressed in these astrophysical units 
is 
\begin{eqnarray}
\dot {m}(x) =   
\nonumber \\
0.1910\,\mu\,{\it n_0}\,{\it x_1}\, 
\left( \tan \left( 0.008725\,{\it 
\alpha_{deg}} \right) \right) ^{3}{\it u_{100}}
{\it d_1}\,
\nonumber \\
\frac { {\mathcal {M}}_{\sun}} {\mbox {year4}} 
, 
\label{mass_astro}
\end{eqnarray}
where $n_0$ is the
number density expressed in particles~$\mathrm{cm}^{-3}$~
(density~$\rho=n_0$m, where $m=\mu m_{\mathrm {H}}$) ,
$\mu$ is the mean molecular weight
(see \cite{Dalgarno1987} suggests $\mu$=1.4~)
, 
$m_{\mathrm {H}}$ is the hydrogen mass
, ${\mathcal {M}}_{\sun}$ is the mass of the sun and
$yr4$ are $10^4~year$.
On introducing the solar system abundances , $N(El)$ ,
where $El$ represents the considered element 
, see Table~2 in \cite{Lodders2003} , 
and the time expressed in $year$ we
obtain 
\begin{eqnarray}
\dot {m}(x) =
\nonumber  \\
1.9\times 10^{-5}\,\mu\,{\it n_0}\,{\it x_1}\, 
\left( \tan \left( 0.008725\,{\it 
\alpha_{deg}} \right) \right) ^{3}{\it u_{100}}\,
{\it d_1}
\nonumber  \\
\frac{N(El)}{N(H)} \quad \frac { {\mathcal {M}}_{\sun}} {\mbox {year}} 
\quad ,
\label{mass_astro_el}
\end{eqnarray}
where $N(H)$ is the Hydrogen solar system abundance. 

As an example when the $Fe$ is considered we obtain
\begin{eqnarray}
\dot {m}(x) =
\nonumber \\
6.59\times 10^{-10}\,\mu\,{\it n_0}\,{\it x_1}\, 
\left( \tan \left( 0.008725\,{\it 
\alpha_{deg}} \right) \right) ^{3}{\it u_{100}}\,
{\it d_1}
\nonumber  \\
\quad \frac { {\mathcal {M}}_{\sun}} {\mbox {year}} 
\quad , \\
case~of~Fe \nonumber 
\label{mass_astro_fe}
\end{eqnarray}
where $\frac{N(Fe)}{N(H)}= \frac {8.380 \times 10^{5}}
{2.431 \times 10^{10}}$ 
,see Table~2 in \cite{Lodders2003}.

A comparison of the previous formula can be done
with ${\dot {\mathcal M}}([FeII])$ that in
HH1 varies between 
$2.2 \times 10^{-7} {\mathcal {M}}_{\sun} yr^{-1}$
and 
$2.8 \times 10^{-9} {\mathcal {M}}_{\sun} yr^{-1}$
, see Table 3 in \cite{Nisini2005}.

The flow of energy ,equation~(\ref{energy_practical}), 
in these astrophysical units
is
\begin{eqnarray}
\dot {E}(x)
=  \nonumber \\
\frac
{
{ 1.203\times 10^{34}}\,\mu\,{\it n_0}\, \left( \tan \left( 0.00872\,
{\it \alpha_{deg}} \right) \right) ^{5}
{{\it u_{100}}}^{3}{{\it d_1}}^{3}
}
{
{\it x_1}
}
\nonumber \\
\quad 
\frac{ ergs}{s}
\label{luminosity_astro}
\quad . 
\end{eqnarray}
The analysis makes extensive use of Favre's (1969) statistical
mass-averaging technique for compressible turbulent flow

\subsection{The simple solution}

Equation~(\ref{velocita_simple}) allows us 
to deduce the centerline velocity
in the simple case 
\begin{eqnarray}
u_0(x) =   \nonumber \\
\frac
{
 1.021{\it u_{100} }{\it d_1}
}
{
{\it d_1}+ 2\tan \left( 0.00872 {\it \alpha_{deg}} \right) {\it x_1}-
 {\it d_1}\tan \left( 0.00872{\it \alpha_{deg}} \right) 
}
\nonumber \\
\frac{Km}{sec}
\quad . 
\label{u_astro_simple}
\end{eqnarray}

The astrophysical version
of the equation of motion,formula~(\ref{traiettoria_simple}),
is 
\begin{eqnarray}
x(t) =  
 0.5 {\it d_1}- 0.5 {\frac {{\it d_1}}{\tan \left(  0.00872 {
\it \alpha_{deg}} \right) }}+ 
\nonumber  \\
0.707{\frac {\sqrt { 0.5 {{\it d_1}}^{2
}+
 2.042 {\it d_1} \tan \left(  0.00872 {\it \alpha_{deg}} \right) {\it 
u_{100}} {\it t_4}}}{\tan \left(  0.00872 {\it \alpha_{deg}} \right) }}
 .
\label{x_astro_simple}
\end{eqnarray}

The flow rate of mass , see equation~(\ref{flowmass_simple}) ,
is 
\begin{eqnarray}
\dot {m}(x) = 
0.01259\, ( 0.5\,{\it d_1}+\tan \left( 0.00872\,{\it 
\alpha_{deg}} \right) {\it x_1}
\nonumber \\ 
- 0.5\,{\it d_1}\,\tan \left( 0.00872\,{
\it \alpha_{deg}} \right) ) {\it d_1}\,{\it u_{100}}\,\mu\,{\it n_0}\,{
}
\frac { {\mathcal {M}}_{\sun}} {\mbox {year4}} 
\quad .
\label{mass_astro_simple}
\end{eqnarray}

The flow of kinetic energy ,equation~(\ref{flowenergy_simple}), 
is
\begin{equation}
\dot {E}(x) 
=
\frac
{
{ 6.231\times 10^{33}}\,\mu\,{\it n_0}\,{{\it u_{100}}}^{3}{{\it d_1}}^{3}
}
{
{\it d_1}+ 2.0\,\tan \left( 0.00872\,{\it \alpha_{deg}} \right) {\it x_1}
}
\quad 
\frac{ ergs}{s}
\label{luminosity_astro_simple}
\quad . 
\end{equation}

\section{Complex trajectories}

\label{sec_complex}
This section reports the kinematic effects that 
lead to complicate trajectories as well an explanation
for the train of knots which are visible in the first part
of the HH objects.

\subsection{The precessing jets}
\label{precessing}
The wide spectrum of observed morphologies that 
characterizes the
HH objects can be due to the kinematic
effects as given by the composition of the velocities of 
different effects such as 
decreasing jet velocity ,
jet precession and proper velocity of the host 
star in the interstellar medium (ISM).
Of particular
interest is the evaluation of various matrices that will
enable us to cause a  transformation from the inertial coordinate
system of the jet to the coordinate system in which the host
star is moving in space. 
The various coordinate systems will be
${\bf x}$=$(x,y,z$) , ${\bf x}^{(1)}$=$(x^{(1)},y^{(1)},z^{(1)})$
, $\ldots$ ${\bf x}^{(3)}$=$(x^{(3)},y^{(3)},z^{(3)})$. 
The
vector representing the motion of the jet is 
represented by
the following $1 \times 3$ matrix
\begin{equation}
G=
\left[ \begin {array}{c} x \left( t \right) \\\noalign{\medskip}0
\\\noalign{\medskip}0\end {array} \right] 
\quad ,
\end{equation}
where the jet motion L(t) is considered along the x-axis.

The jet axis, $x$, is inclined at an angle $\Psi_{prec}$
relative to an axis $x^{(1)}$ and therefore
the $3 \times 3$ matrix,
representing a rotation through the z axis,
is given by:
\begin {equation}
F=
 \left[ \begin {array}{ccc} \cos \left( \Psi_{{{\it prec}}} \right) &-
\sin \left( \Psi_{{{\it prec}}} \right) &0\\\noalign{\medskip}\sin
 \left( \Psi_{{{\it prec}}} \right) &\cos \left( \Psi_{{{\it prec}}}
 \right) &0\\\noalign{\medskip}0&0&1\end {array} \right] 
\quad .
\end {equation}
From a practical point of view $\Psi_{prec}$ can be derived by
measuring the half opening angle of the maximum of the sinusoidal
oscillations that characterizes the jet.

\begin{figure}
 \begin{center}
\includegraphics[width=7cm]{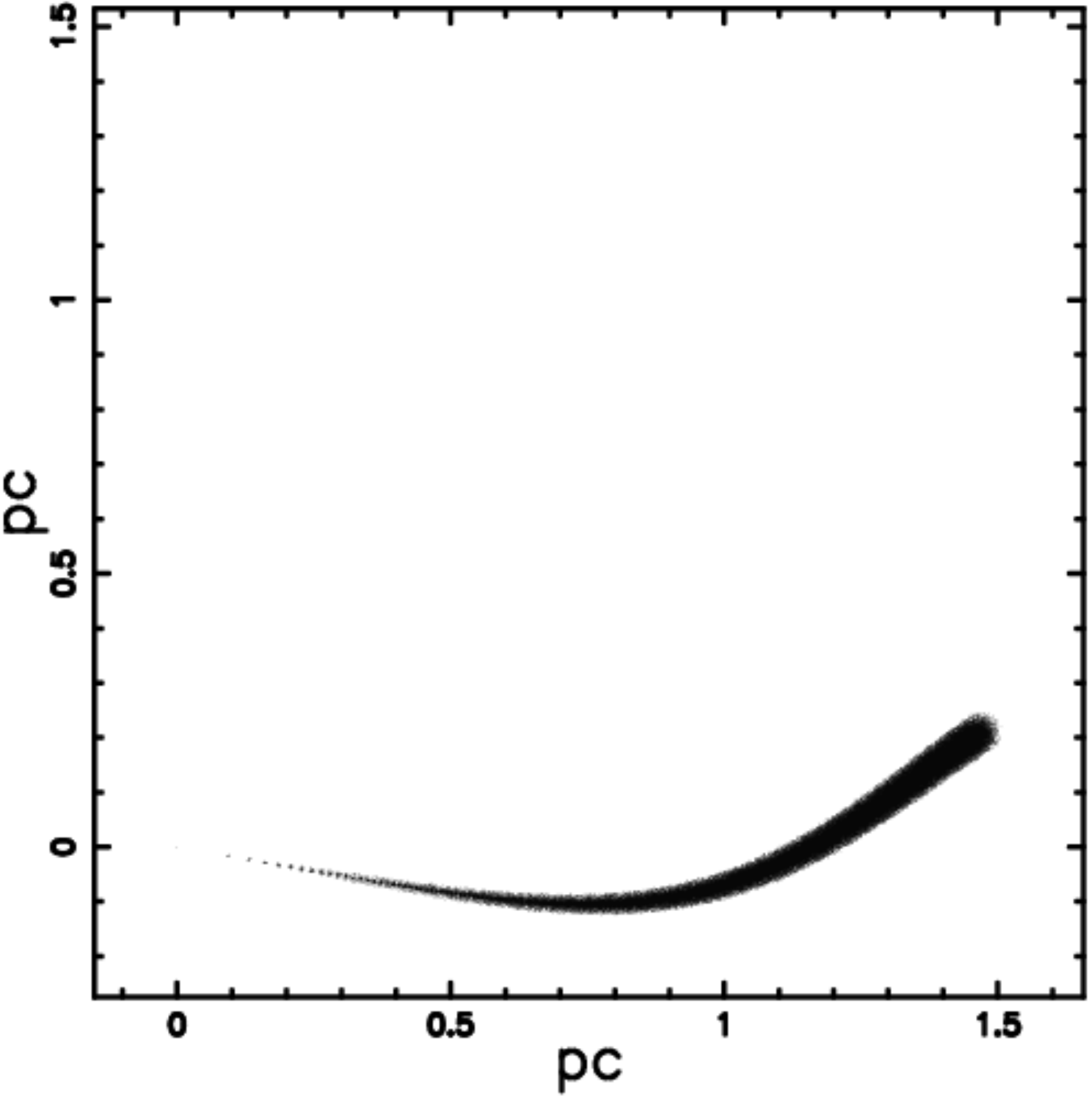}
 \end {center}
\caption {
 Continuous three-dimensional trajectory of HH34:
 the three Eulerian angles 
 characterizing the point of view are 
   $ \Phi $= 0 $^{\circ }$
 , $ \Theta $= 0 $^{\circ }$
 and $ \Psi $= 0 $^{\circ }$.
 The precession is characterized by the angle
 $ \Psi_{prec} $= 10 $^{\circ }$
 and by the angular velocity
 $ \Omega_{prec} $= 36.00 [$^{\circ}/10^4\mathrm{year}$].
 The physical parameters characterizing the 
 jet motion
 are :
 $u_{100}=6.5 $,
 $t_4 = 4 $ ,
 $x_1 = 1.49$ ,
 $d_1 = 9.7 \,10^{-4} $ 
 and 
 $\alpha_0$= 2.86 $^{\circ }$.
}
\label{hh34_proj_1}
\end{figure}

If the jet is undergoing precession around the
$x^{(1)}$ axis, $\Omega_{prec}$ can be
the angular velocity of precession expressed in
$\mathrm{radians}$ per unit time ; 
 $\Omega_{prec}$ is computed from the optical maps
by measuring the
number of sinusoidal oscillations
that characterize the jet.
The transformation from the coordinates
${\bf x}^{(1)}$ fixed in the frame of the
precessing jet to the non-precessing coordinate
${\bf x}^{(2)}$
is represented by the $3 \times 3$ matrix
\begin{equation}
 P=
\left[ \begin {array}{ccc} 1&0&0\\\noalign{\medskip}0&\cos \left( 
\Omega_{{{\it prec}}}t \right) &-\sin \left( \Omega_{{{\it prec}}}t
 \right) \\\noalign{\medskip}0&\sin \left( \Omega_{{{\it prec}}}t
 \right) &\cos \left( \Omega_{{{\it prec}}}t \right) \end {array}
 \right] 
\quad .
\end{equation}
As an example Figure~\ref{hh34_proj_1} reports
the precessing jet applied to HH34.

\begin{figure}
 \begin{center}
\includegraphics[width=7cm]{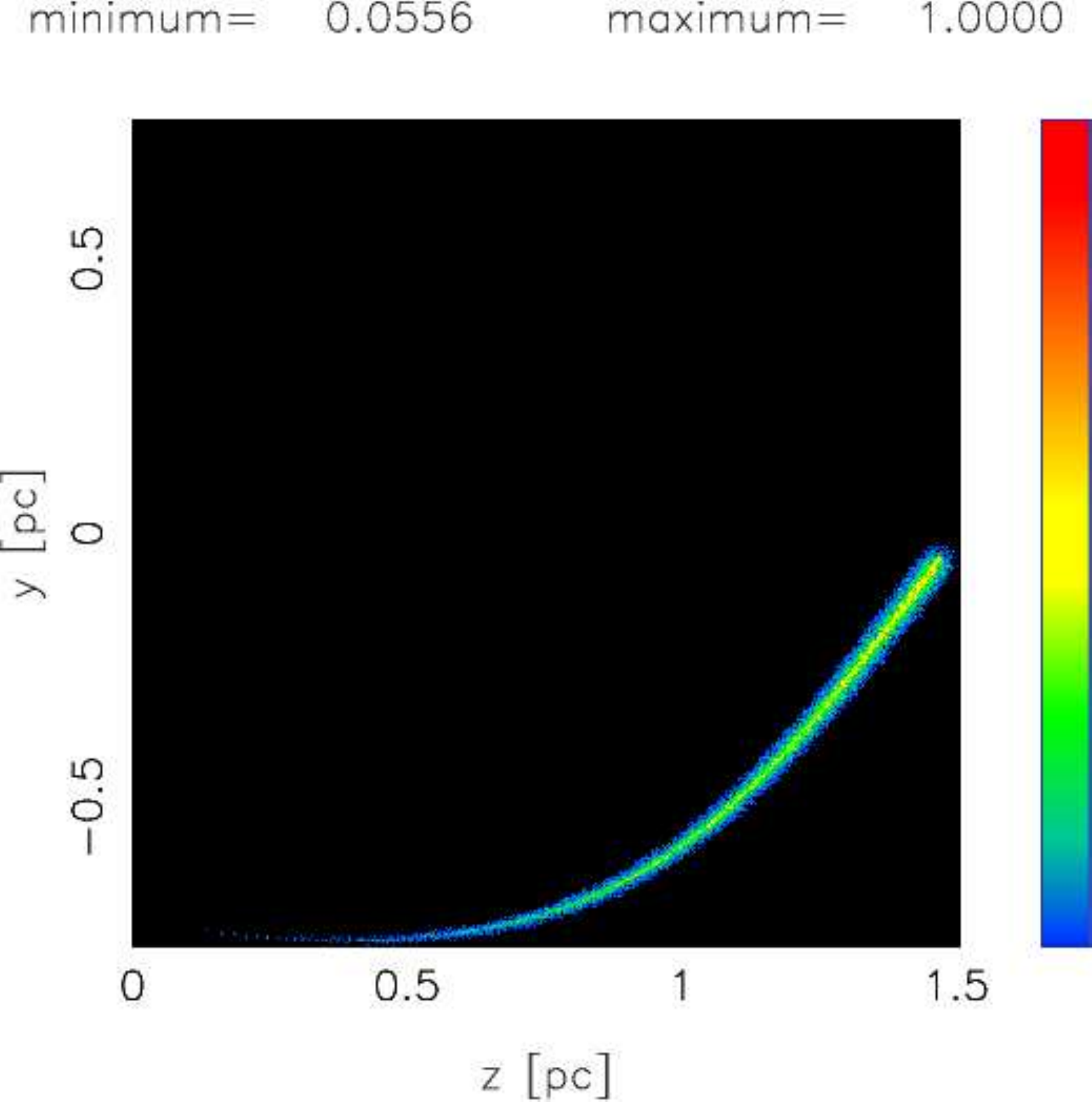}
 \end {center}
\caption {
 Continuous three-dimensional trajectory of HH34:
 the three Eulerian angles 
 characterizing the point of view are 
 $ \Phi $= 0 $^{\circ }$
 , $ \Theta $= 0 $^{\circ }$
 and $ \Psi $= 0 $^{\circ }$.
 The precession is characterized by the angle
 $ \Psi_{prec} $= 10 $^{\circ }$
 and by the angular velocity
 $ \Omega_{prec} $= 36.00 [$^{\circ}/10^4\mathrm{year}$].
 The star has velocity $v_y=11.19 \frac{Km}{s}$.
 The physical parameters characterizing the 
 jet motion
 are :
 $u_{100}=6.5 $,
 $t_4 = 4 $ ,
 $x_1 = 1.49$ ,
 $d_1 = 9.7 \,10^{-4} $ 
 and 
 $\alpha_0$= 2.86 $^{\circ }$.
Here the plane of the trajectory is 
perpendicular to the observer .
The two Cartesian axis are in $pc$ 
and 
the integral operation which allows to 
build the image 
is performed on cubic grid 
of $1200^3$ pixels.
}
\label{hh34_complex_proj_2}
\end{figure}

The last translation represents
the change of framework from
$\bf(x^{(2)})$, which is co-moving with
the host star, to a system
$\bf (x^{(3)})$ in comparison to which the
host star is in  a uniform
motion.
In the laboratory experiments the velocity of 
the host star is replaced by a wind ,
see Figure~3 in \cite{Ciardi2008}.
The relative motion of the origin of the coordinate system
$(x^{(3)},y^{(3)},z^{(3)})$ is defined by the
Cartesian components of the star velocity $ v_x,v_y,v_z$,
and the required $1 \times 3 $ 
matrix transformation representing this translation is:
\begin{equation}
 B=
\left[ \begin {array}{c} v_{{x}}t\\\noalign{\medskip}v_{{y}}t
\\\noalign{\medskip}v_{{z}}t\end {array} \right] 
\label {transla}
\quad .
\end{equation}
On assuming, for the sake 
of simplicity, that $v_x$=0 and $v_z$=0,
the translation matrix becomes:
\begin{equation}
 B=
\left[ \begin {array}{c} 0\\\noalign{\medskip}v_{{y}}t
\\\noalign{\medskip}0\end {array} \right] 
\quad .
\label{traslationmatrix}
\end{equation}
In other words, the direction of the star motion in the ISM and
the direction of the jet are perpendicular. From a practical point
of view the star velocity can be measured by dividing the
length of the star in a direction perpendicular to the
initial jet velocity by the lifetime of the jet.
 The
final $1 \times 3$ matrix $A$ representing the
``motion law'' can be found
by composing the four matrices already
described
\begin {eqnarray}
\lefteqn {A = B + ( P \cdot F \cdot G) } \nonumber \\
 =&
\left[ \begin {array}{c} \cos \left( \Psi_{{{\it prec}}} \right) x
 \left( t \right) \\\noalign{\medskip}v_{{y}}t+\cos \left( \Omega_{{{
\it prec}}}t \right) \sin \left( \Psi_{{{\it prec}}} \right) x \left( 
t \right) \\\noalign{\medskip}\sin \left( \Omega_{{{\it prec}}}t
 \right) \sin \left( \Psi_{{{\it prec}}} \right) x \left( t \right) 
\end {array} \right] 
\quad .
\end {eqnarray}
The three components of the previous
 $1\times 3$ matrix $A$
represent the jet motion
along the Cartesian coordinates as given by the observer
that sees the star moving in a uniform motion.
As an example Figure~\ref{hh34_complex_proj_2} reports
the effect of inserting the star's velocity on 
the precessing HH34 as plotted in Figure~\ref{hh34_proj_1}.

\begin{figure}
 \begin{center}
\includegraphics[width=7cm]{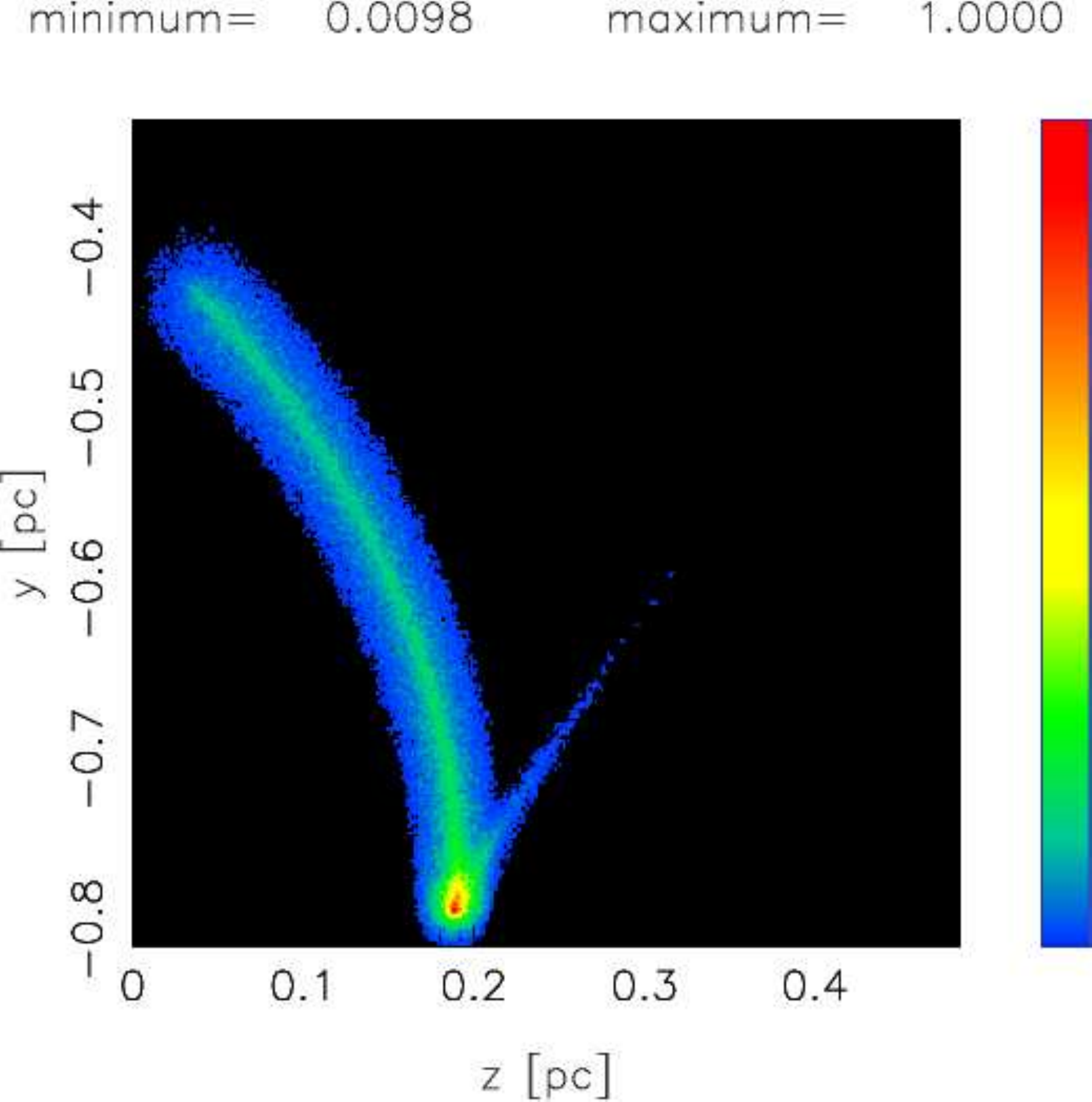}
 \end {center}
\caption {
 Continuous three-dimensional trajectory of HH34:
 the three Eulerian angles 
 characterizing the point of view are 
 $ \Phi $= 290 $^{\circ }$
 , $ \Theta $= 180 $^{\circ }$
 and $ \Psi $= 18 $^{\circ }$.
 The precession is characterized by the angle
 $ \Psi_{prec} $= 10 $^{\circ }$
 and by the angular velocity
 $ \Omega_{prec} $= 36.00 [$^{\circ}/10^4\mathrm{year}$].
 The star has velocity $v_y=11.19 \frac{Km}{s}$.
 The physical parameters characterizing the 
 jet motion
 are :
 $u_{100}=6.5 $,
 $t_4 = 4 $ ,
 $x_1 = 1.49$ ,
 $d_1 = 9.7 \,10^{-4} $ 
 and 
 $\alpha_0$= 2.86 $^{\circ }$.
The image is here reported as a theoretical 2D 
surface  brightness  of emission, 
the two Cartesian axis are in $pc$ 
and 
the integral operation is performed on cubic grid 
of $1200^3$ pixels.
}
\label{hh34_complex_zoom}
\end{figure}

The fifth matrix allows to model the point of view 
of the observer through  
the matrix $E$ representing 
the three Eulerian angles 
which characterizes the point of view of the observer,
$\Theta,\Phi,\Psi$ , see \cite{Goldstein2002} .
The product $ E \cdot A$ is not reported 
for space problem and
Figure~\ref{hh34_complex_zoom} reports the same as 
Figure~\ref{hh34_complex_proj_2} , 
but from a particular point of view.
In other words the particular point of view
can produce complex projected patterns
of a simple basic trajectory 
as represented by Figure~\ref{hh34_complex_proj_2}.
A comparison of Figure 
\ref{hh34_complex_zoom} should be done with 
the image of HH34 as available at               \\
http://antwrp.gsfc.nasa.gov/apod/ap991129.html  \\
made with the VLT by the FORS Team or
Figure 1 in \cite{Reipurth2002} 
which has a field 
of $4.5^{\prime} \times 5^{\prime}$.
The astrophysical version 
of the star's motion as represented
by  the translation matrix $B$, 
formula~(\ref{traslationmatrix}),
is 
\begin{equation}
y=0.01021\,{\it v_y}\,{\it t_4}
\quad ,
\end{equation}
where 
$v_y$ is expressed in $\frac{Km}{s}$ units 
and 
$t_4$ = $t[yr]/(10^4)$ .

The previous equation
can be combined with the motion 
 along 
$x$ as represented  
by equation~(\ref{x_astro}) in order
to find the angle $\beta $ in degree 
that characterizes 
the  trajectory:
\begin{eqnarray}
\beta=\arctan (\frac {y}{x})  = \nonumber \\
57.29 \arctan ( \frac{
0.009718  v_y \sqrt {{\it t_4}}
}
{
\sqrt {\it u_{100}}  \sqrt { d_1 }
\sqrt {  \tan \left(  0.008727
{\it \alpha_{deg} } \right)}
} )
~~^\circ .
\end{eqnarray}
This angle varies from 0 when $t_4$=0 to 
$23.39^\circ$ when $t_4$=4  and the parameters   
of Figure (\ref{hh34_complex_proj_2}) are used.

Is also interesting to point out that 
a rotation of $90^{\circ}$ around the $y$ axis 
of the trajectory as reported in
Figure~\ref{hh34_complex_proj_2}  makes the jet
straight  rather than bended.

Analogous results on ballistic jets
from precessing sources has been obtained, see 
\cite{Lightfoot1986,Raga1993}.

\subsection{The Kelvin-Helmholtz instabilities }

The macroscopic phenomena of the jets as the 
presence of knots and wiggles 
can be due to the 
Kelvin-Helmholtz instability 
(after \cite{Kelvin,Helmholtz}) 
of an 
axisymmetric flow along the
velocity-axis when the 
wavelengths $\lambda=\frac{2 \pi}{k} $ ( $k$ is the
wave-vector) are greater than the jet radius $a_j$,
which is taken to be independent of the position 
along the jet, see  
\cite{Zaninetti_1979,Zaninetti1981,Ray1983,hardee}.
The velocity , $U_0$, is assumed to be rectangular. 
The internal (
external ) fluid density is represented by $\rho_{0i}$
($\rho_{0e}$) , the internal sound velocity is $s_i$ 
and $\nu_0$=
$\frac {\rho_{0i}}{\rho_{0e}}$. 
Starting from the equations of motion
and continuity, and
assuming both fluids to be adiabatically compressible,
it is possible to derive and to solve
the dispersion relation from a numerical 
point of view  , see \cite{zaninetti1987}.

We then start from observable quantities
that can be measured on radio-maps
such as the total length $L_{obs}$,
the wavelength $\lambda_{obs}^1$ of the wiggles ($m$=1)
along the jet,
the distance $\lambda_{obs}^0$ ($m$=0) between knots,
and the final offset
$\Delta\,L_{obs}$ of the center of the jet.

These observable quantities are identified
with the following theoretical variables:
\begin {equation}
 \lambda^1_{max} = \lambda^1_{obs}
\quad,
\end {equation}
\begin {equation}
 \lambda^0_{max} = \lambda^0_{obs}
\quad,
\end {equation}
\begin {equation}
 A_0 \exp \bigl (\frac {2L_{obs}} {M t_{ad} a_j}) = \Delta L_{obs}
\quad,
\end {equation}
\begin {equation}
 nl_e = L_{obs}
\quad,
\end {equation}
where $t_{ad}$ = $t_{min } \cdot s_i/a_j $
and $A_0$ is the amplitude of the
perturbed energy.
The result is a theoretical expression for $t_{min}$
the minimum time scale of the instability,
$\lambda_{max}$ the wavelength connected with the most
unstable mode and $l_e$ the distance over which the most
unstable mode grows by a factor $e$, 
see
\cite{zaninetti1987}.
These parameters can then be found
through the set of nonlinear equations
previously reported.
By choosing two objects,
HH1
and
HH34 
the observational parameters can be measured
on the optical image, see Table~\ref{observ}.
 \begin{table}
 \caption{Parameters of the observed oscillations in HH objects
expressed in averaged radius units }
 \label{observ}
 \[
 \begin{array}{lcc}
 \hline
 \hline
 \noalign{\smallskip}
Geometrical~measure & HH1 & HH34 \\
 \noalign{\smallskip}
 \hline
 \noalign{\smallskip}
L_{obs} [averaged~radius~units] & 73.3 & 54 \\ \noalign{\smallskip}
\lambda^0_{obs}[averaged~radius~units] & 8.1 & 4.5 \\ \noalign{\smallskip}
\lambda^1_{obs}[averaged~radius~units] & 29.3 & 27 \\ \noalign{\smallskip}
\Delta L_{obs} [averaged~radius~units] & 1 & 1 \\ \noalign{\smallskip}
\noalign{\smallskip}
 \hline
 \hline
 \end{array}
 \]
 \end {table}

The four nonlinear equations are then solved
and the four theoretical parameters
are found , see Table~\ref{theor}.
 \begin{table}
 \caption{Theoretical parameters from oscillations deduced
from the four nonlinear equations }
 \label{theor}
 \[
 \begin{array}{lcc}
 \hline
 \hline
 \noalign{\smallskip}
Theoretical~variable & HH1 & HH34 \\
 \noalign{\smallskip}
 \hline
 \noalign{\smallskip}
n & 1.22 & 1.96 \\ \noalign{\smallskip}
M & 52.2 & 26.67 \\ \noalign{\smallskip}
\nu_0 & 325 & 15.67 \\ \noalign{\smallskip}
A_0 & 0.012 & 0.017 \\ \noalign{\smallskip}
 \hline
 \hline
 \end{array}
 \]
 \end {table}

\begin{figure}
 \begin{center}
\includegraphics[width=7cm]{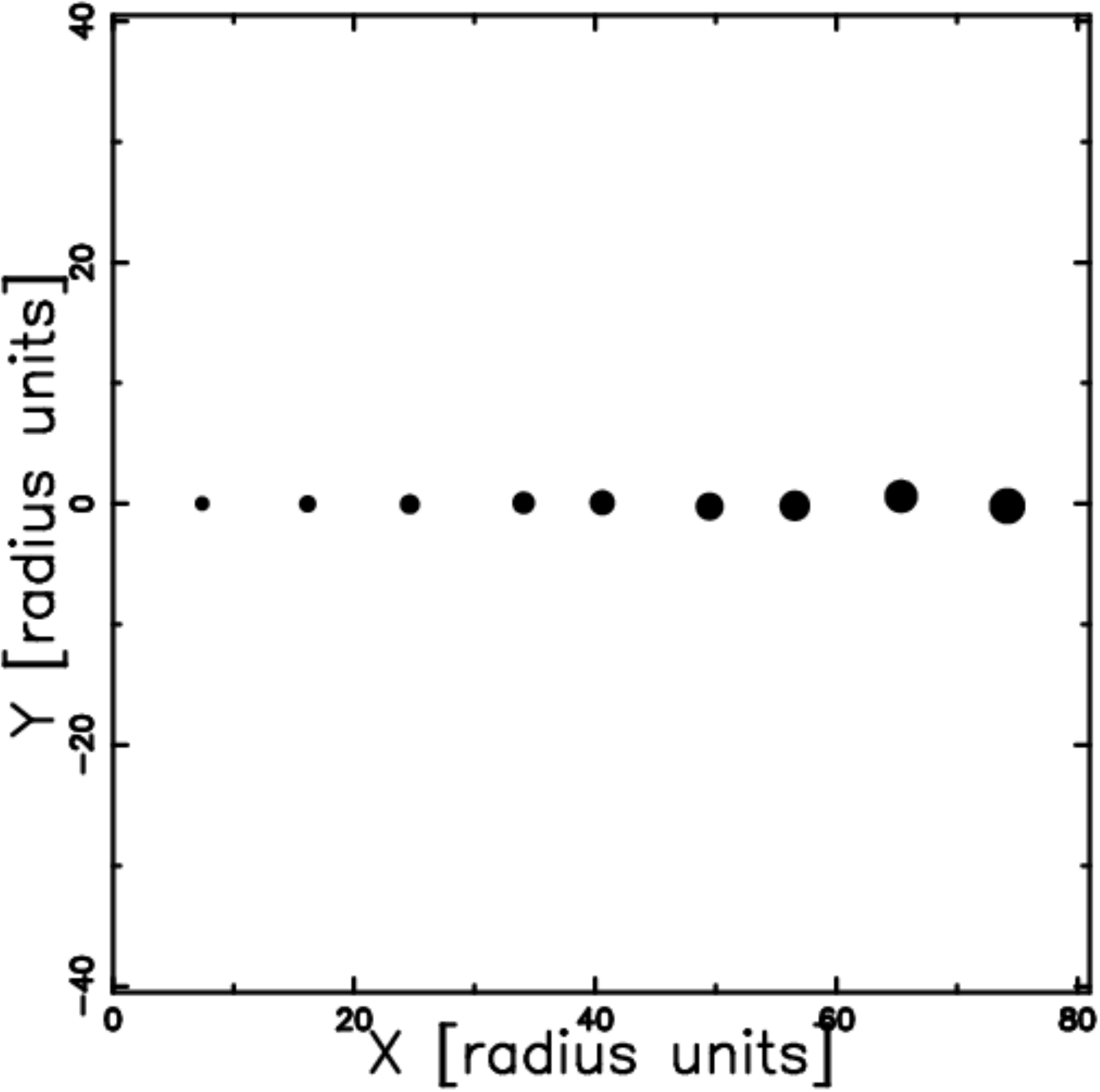}
 \end {center}
\caption {
 Superposition of the pinch mode ($m=0$)
 on the helical mode ($m=1$) for HH1 in radius
 units.
 The parameters are as in Tables \ref{observ} and \ref{theor}.
}
\label{hh1_kh}
\end{figure}

\begin{figure}
 \begin{center}
\includegraphics[width=7cm]{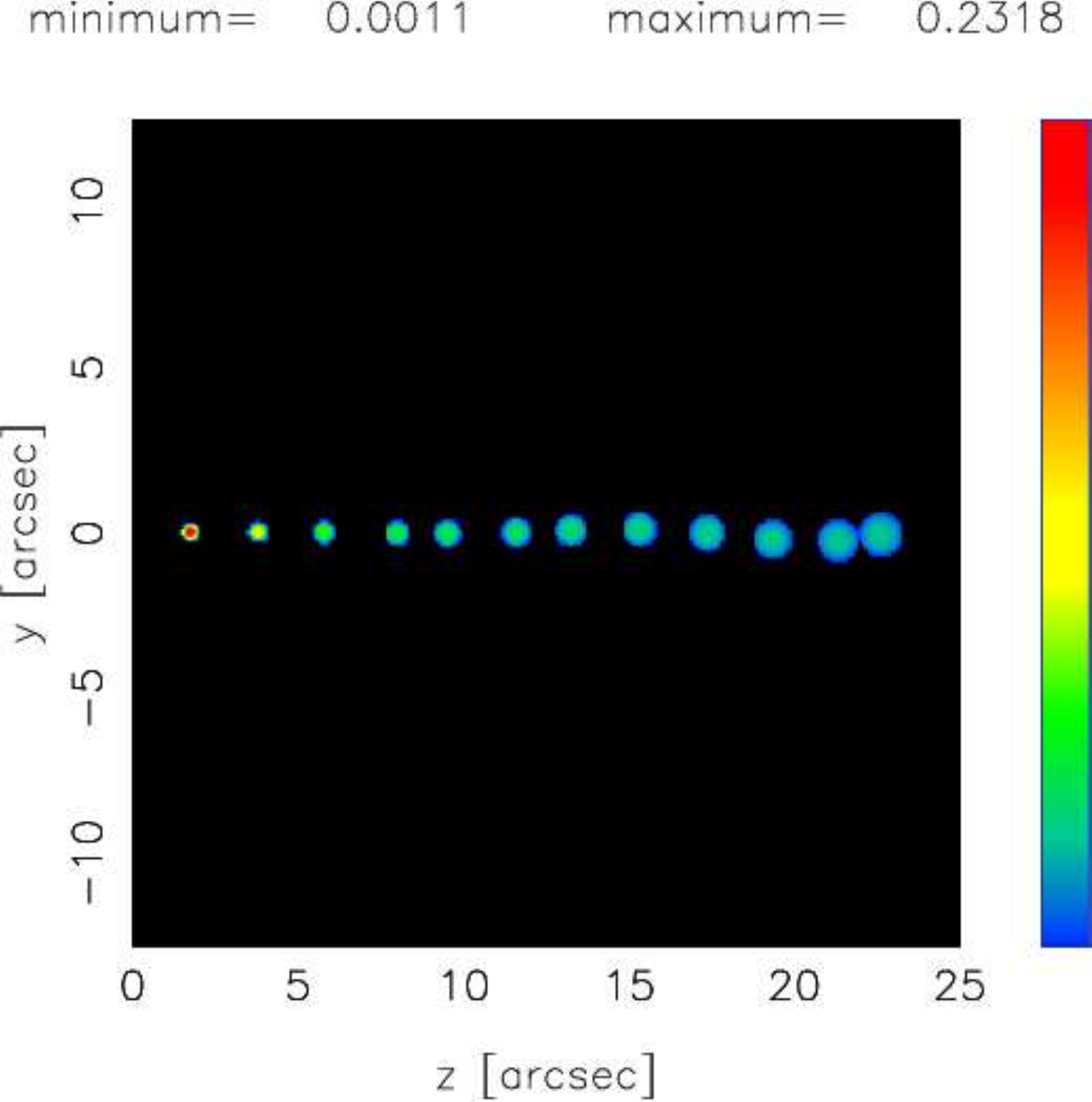}
 \end {center}
\caption {
 Superposition of the pinch mode ($m=0$)
 on the helical mode ($m=1$) for HH34 in radius
 units.
 The parameters are as in Tables \ref{observ} and \ref{theor}.
The image is here reported as a theoretical 2D 
surface brightness  of emission, 
the integral operation is performed on cubic grid 
of $1200^3$ pixels.
The  intensity or brightness  in the  spherical blobs ($m=0$) 
scales as in 
equation (\ref{icylindera}).
}
\label{hh34int}
\end{figure}

An application of the results 
for  HH1 
here obtained is reported
in Figure~\ref{hh1_kh} ; 
the comparison should be done with Figure 1 ($[SII]$)
in \cite{Reipurth2000} that covers 
$\approx$ 14.16 arcseconds.
The application to HH34 is
reported in Figure~\ref{hh34int}
and the comparison should be done with 
Figure 3 in \cite{Reipurth2002} which covers 
$\approx 30$ arcsec.
In both cases  the wavelength
of the pinch modes ($m=0$)
and the oscillations of the helical mode ($m=1$)are those
reported in Table~\ref{theor}.

\section{The image from geometry }

\label{sec_image}
The transfer equation in the presence of emission only
, see for example equation~(1.27)  in   
\cite{rybicki}
or equation~(4.9) in \cite{Dopita2003} 
 ,
 is
 \begin{equation}
\frac {dI_{\nu}}{ds} =  -k_{\nu} \zeta I_{\nu}  + j_{\nu} \zeta
\label{equazionetrasfer}
\quad ,
\end {equation}
where  $I_{\nu}$ is the specific intensity
or  brightness
which has units of $\frac {erg}{s\, cm^2\, ster \,Hz}$ 
 , $s$  is the
line of sight , $j_{\nu}$ the emission coefficient 
which has units of $\frac {erg}{s\, cm^3\, ster \,Hz}$ ,
$k_{\nu}$   a mass absorption coefficient,
$\zeta$ the  mass density at position s
and the index $\nu$ denotes the interested frequency of
emission.
The solution to  equation~(\ref{equazionetrasfer})
 is
\begin{equation}
 I_{\nu} (\tau_{\nu}) =
\frac {j_{\nu}}{k_{\nu}} ( 1 - e ^{-\tau_{\nu}(s)} )
\quad  ,
\label{eqn_transfer}
\end {equation}
where $\tau_{\nu}$ is the optical depth at frequency $\nu$
\begin{equation}
d \tau_{\nu} = k_{\nu} \zeta ds
\quad.
\end {equation}
We now continue analyzing the case of
an
 optically thin layer
in which $\tau_{\nu}$ is very small
( or $k_{\nu}$  very small )
and the density  $\zeta$ is substituted
with our number density C(s) of  particles.
Two cases are taken into account :   
the  emissivity is proportional
to the number density and the emissivity is   
proportional to the square of the number density .
In the  linear case 
\begin{equation}
j_{\nu} \zeta =K  C(s)
\quad  ,
\end{equation}
where $K$ is a  constant function.

In the  quadratic  case 
\begin{equation}
j_{\nu} \zeta =K_2  C(s)^2
\quad  ,
\label{eqn_transfer_square}
\end{equation}
where $K_2$ is a  constant function.
This is true for example for  
free-free radiation from a thermal plasma,
see formula (1.219) in  \cite{lang} or
formula (6.17) in \cite{Dopita2003} .

The intensity is now
\begin{eqnarray}
 I_{\nu} (s) = K
\int_{s_0}^s   C (s\prime) ds\prime \\
 \mbox {optically thin layer}
\quad linear~case \quad ,           \nonumber 
\label{transport1}
\end {eqnarray}
or 
\begin{eqnarray}
 I_{\nu} (s) = K_2 
\int_{s_0}^s   C (s\prime)^2 ds\prime \\
\quad  \mbox {optically thin layer}
\quad quadratic~case \quad . \nonumber
\label{transport2}
\end {eqnarray}
In the Monte Carlo experiments
the number density is memorized  on
a 3D   grid
${\mathcal M(i,j,k)}$ 
where $i,j$ and $k$ are indexes 
varying from 1 to $pixels$ , 
and the intensity is
\begin{eqnarray}
{\it I}\/(i,j) = \sum_k  \triangle\,s \times  {\mathcal M}(i,j,k)
\\
\quad  \mbox {optically thin layer}
\quad linear~case 
\quad,   \nonumber 
\label{thin1}
\end{eqnarray}
or 
\begin{eqnarray}
{\it I}\/(i,j) = \sum_k  \triangle\,s \times  {\mathcal M}(i,j,k)^2 
\\
\quad  \mbox {optically thin layer}
\quad quadratic~case 
\quad  ,   \nonumber
\label{thin2}
\end{eqnarray}
where $\triangle$s is the spatial interval between
the various values of intensity 
and  the sum is performed
over the   interval of existence of the index $k$.
In this grid framework 
the little squares that
characterized by the position 
of the indexes $i,j$  correspond to a
different line of sight.
When all the different pixels are viewed together
the image is formed.
The ensemble of all the pixels can be considered
a theoretical surface intensity 
or a theoretical surface brightness.
We now outline a possible source of radiation.
The volume emission coefficient of 
the transition $j_{21} $ 
is 
\begin{equation}
j_{21} = \frac{n_2 A_{21} h \nu_{21} } { 4 \pi}
\quad  ,
\end{equation}
where level 1 is  the lower level ,  
level 2 is  the upper level       ,
$n_2 $ is gas number density ,
$n_2 A_{21}$ the rate of photons emitted from 
a unit volume ,
$A_{21}$ is the Einstein coefficient for the transition ,
$h$  is the Planck constant and 
$\nu_{21}$ the considered frequency,
see  \cite{Hartigan2008}.
In the case of  optically thin medium 
the intensity of the emission $I_{21} $
is the integral along the line of sight 
\begin{equation}
I_{21} = \int j_{21} dl   
\quad .
\end {equation}
In the case of constant gas number density
\begin{equation}
I_{21} \propto l   
\quad ,
\end {equation}
where $l$ is the considered length that 
in the astrophysical diffuse objects 
depends 
from the point of view of the observer.
The optically thin layer approximation represents 
therefore a useful approximation to build models 
for the intensity of radiation which are comparable
to the observed profiles.

We now  analyze the behavior  of the intensity
of a cross section of a jet ,
the behavior of the maximum intensity at the centerline
as a function of the distance from the central source ,
the intensity of complex morphologies 
and the sudden increase in intensity 
as given by the toroidal jet.

\subsection{Intensity at a fixed distance} 

\label{spherical}

\begin{figure}
 \begin{center}
\includegraphics[width=7cm]{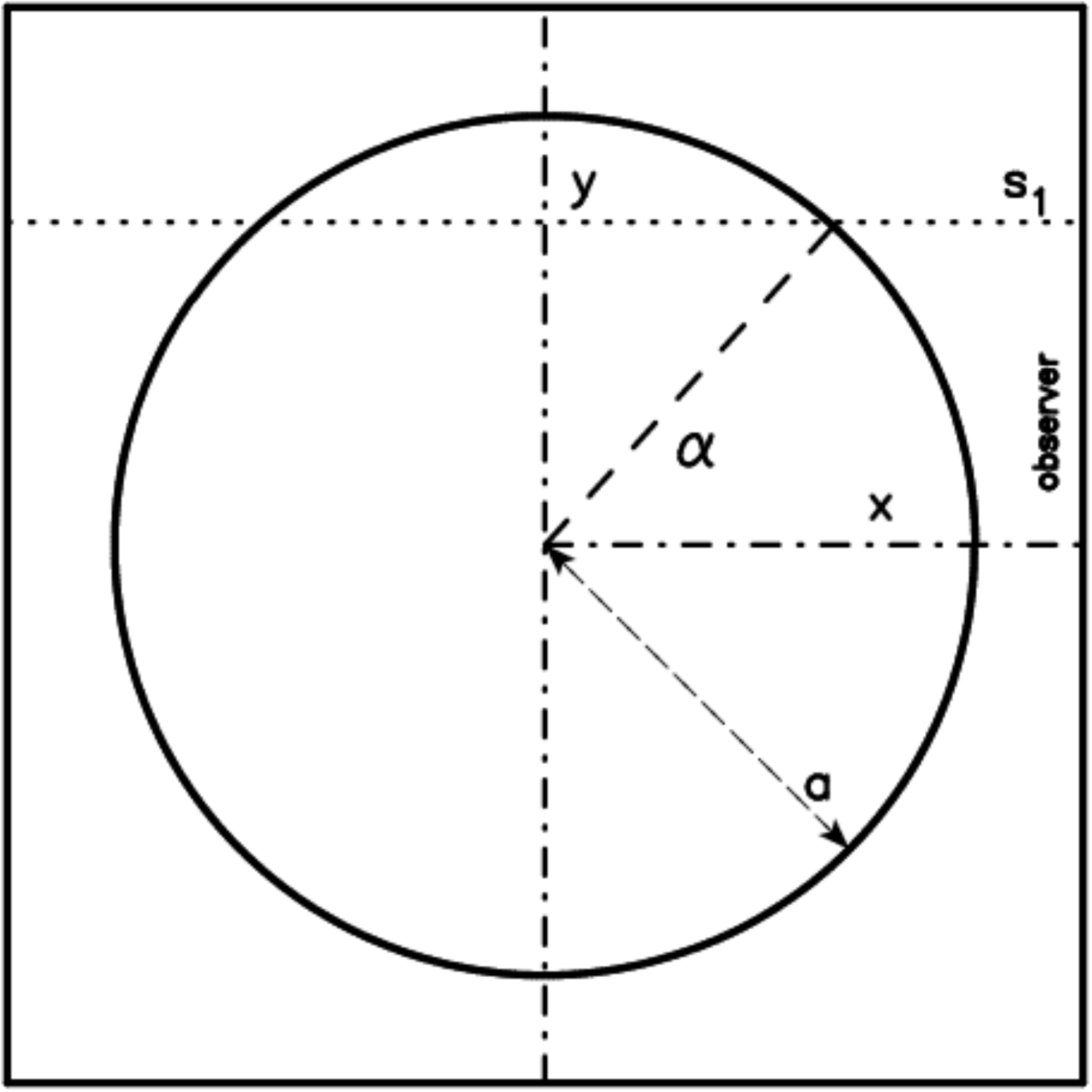}
 \end {center}
\caption {
The source is represented through
a circular section perpendicular to the jet axis.
The observer is situated along the x direction,
one line of sight is indicated
and the angle $\alpha$ is clearly indicated.
 }%
 \label{asolo}
 \end{figure}

We explore the behavior of the  intensity or brightness  
along the a jet when the distance from the origin ,$y$,
is fixed .
We assume that the number density $C$ is constant
in a cross section of radius $a$ 
and then falls to 0 , see Figure~\ref{asolo}.

The length of sight , when the observer is situated
at the infinity of the $x$-axis , 
is the locus 
parallel to the $x$-axis which crosses the position $y$ in a 
Cartesian $x-y$ plane and terminates at the external circle
of radius $a$.
The locus length is 
\begin{eqnarray}
l_{ab} = 2 \times ( \sqrt {a^2 -y^2}) 
\quad ; 0 \leq y < a \quad .
\label{lengthcylinder}
\end{eqnarray}
When the number density $C_m$ is constant in the cylinder 
of radius $a$ 
the  intensity or brightness  of radiation is 
\begin{eqnarray}
I_{0a} =C_m \times 2 \times ( \sqrt { a^2 -y^2}) 
 \quad ; 0 \leq y < a \quad ,
\label{icylinder}
\end{eqnarray}
or
\begin{eqnarray}
I_{0a} =C_m \times 2 \times a \times \cos (\alpha) 
 \quad ;-\frac{\pi}{2} \leq \alpha \leq \frac{\pi}{2} \quad,
\label{icylindera}
\end{eqnarray}
that can be named the "trigonometrical law" for the  intensity or brightness .
Is interesting to underline that the two previous 
equations hold for a cylindrical and a conical 
jet as well for a spherical blob when the 
number density is constant.

\subsection{Centerline Intensity function of the distance} 

\begin{figure}
 \begin{center}
\includegraphics[width=7cm]{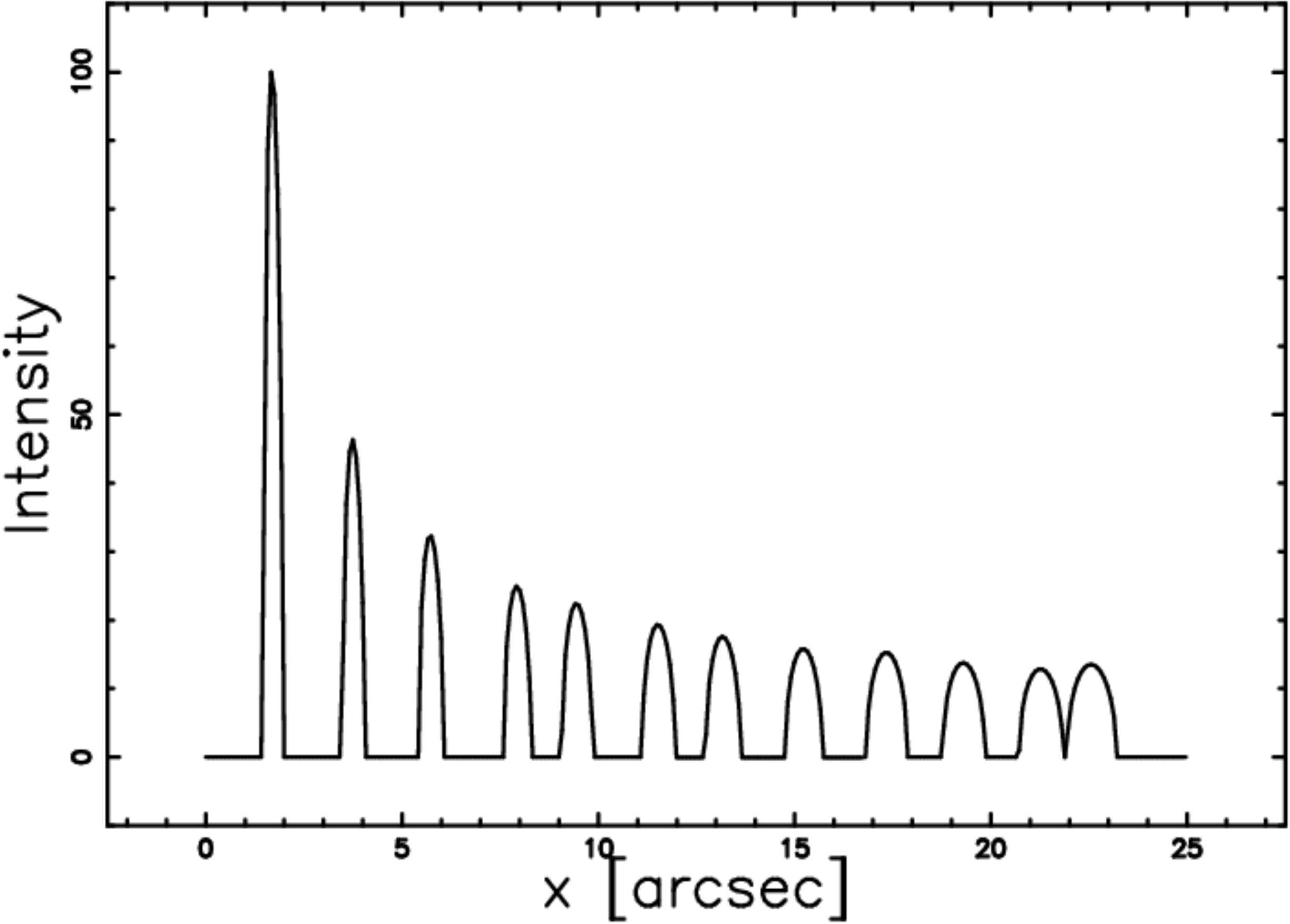}
 \end {center}
\caption {
Intensity of HH34 represented
through a cut in the x-direction (y=0) , parameters as 
in Figure~\ref{hh34int}. 
The  intensity or brightness  along the cut decreases 
as $  I(x)  \propto x^{-0.15}$.
 }%
 \label{cuts_x_hh34}
 \end{figure}

We now explore the behavior of the  intensity or brightness  at 
the centerline of the jet as a function of the distance
$x$ from the nozzle.
From the previous paragraph~\ref{spherical}
we learned that 
the maximum  intensity or brightness  at the centerline
of the jet at a fixed distance $x$ is proportional
, as a first parameter ,
to the jet's diameter $d$ ,
\begin{equation}
I(r=0,x) \propto d(x) C_m \propto x C_m 
\quad .
\end{equation}
In order to have a constant  intensity or brightness  along the 
centerline of the jet as function of $x$ ,
the number density $C_m$ 
of the emitting particles 
should decreases as 
\begin{equation}
C_m(x) \propto \frac{1}{x}
\quad .
\end{equation}
As a consequence the  intensity or brightness  
\begin {equation}
I (x) \propto C_m \times d \propto \frac {cost}{x} x 
\quad ,
\end{equation}
will be constant along the jet.
In the framework of the optically thin medium 
the emitting length will not change 
but conversely the number density
can take the general form 
\begin{equation}
C_m(x) \propto \frac{1}{x^{s+1}}
\quad ,
\end{equation}
which means that the  intensity or brightness  
scales as 
\begin{equation}
I(x) \propto d(x) \frac{1}{x^{s+1}} \propto x^{-s}
\quad .
\label{ilongitudinal}
\end{equation}
The value of $s$ can found from 
the scaling of the observed  intensity or brightness  
as function of $x$.
As an example of constant  intensity or 
brightness  of emission
along a knotty jet we report the image
of the first part of HH34
where  $C_m(x) \propto \frac{1}{x^{1+0.15}}$ was chosen ;
a comparison should be done with Figure~3
in \cite{Reipurth2002}.
The relative cut 
along the jet's main axis of symmetry,
is reported in  Figure~\ref{cuts_x_hh34}
where $s=0.15$ was used.
As a practical example the intensity (counts)
of $H_{\alpha}$ for HH110 as in  Figure 3 
of \cite{Riera2003} decreases of a factor
$\approx $ 8.6 from the second blob B 
to the last blob Q.
In our simulation of HH34 as reported
in Figure \ref{cuts_x_hh34} the intensity 
decreases of 
a factor $\approx$ 8.3 from the first to the last blob.

The 3D algorithm already presented replaces 
previous efforts based on the 2D random walk 
, see Figure 7 in \cite{Zaninetti_1989} and 
on the 3D random walk from many injection points ,
see Figure 8 in \cite{Zaninetti_1999}.

\subsection{Complex Morphologies}

The integral operation of the emissivity along the line 
of sight of a turbulent jet can be performed in an analytical way 
only in a simple configuration : the jet perpendicular
to the observer, see Section~\ref{linear}.
The concurrency of complex trajectories and a general
point of view of the observer
characterized by the three Eulerian angles 
$\Theta$, $\Phi$ and $\Psi$, asks  a numerical treatment.
We remember that the points that characterize the trajectory
of HH34 , see Section~\ref{precessing} , are already 
in such a way that the product 
$\epsilon(x) \times x$ is nearly constant.
This means that the  intensity or brightness  
is nearly constant 
along the main direction.
These points are inserted on a 3D grid made by
$pixels^3$ points and a sum is performed 
over one index, see Figure~\ref{hh34_complex_zoom}. 

\begin{figure}
 \begin{center}
\includegraphics[width=7cm]{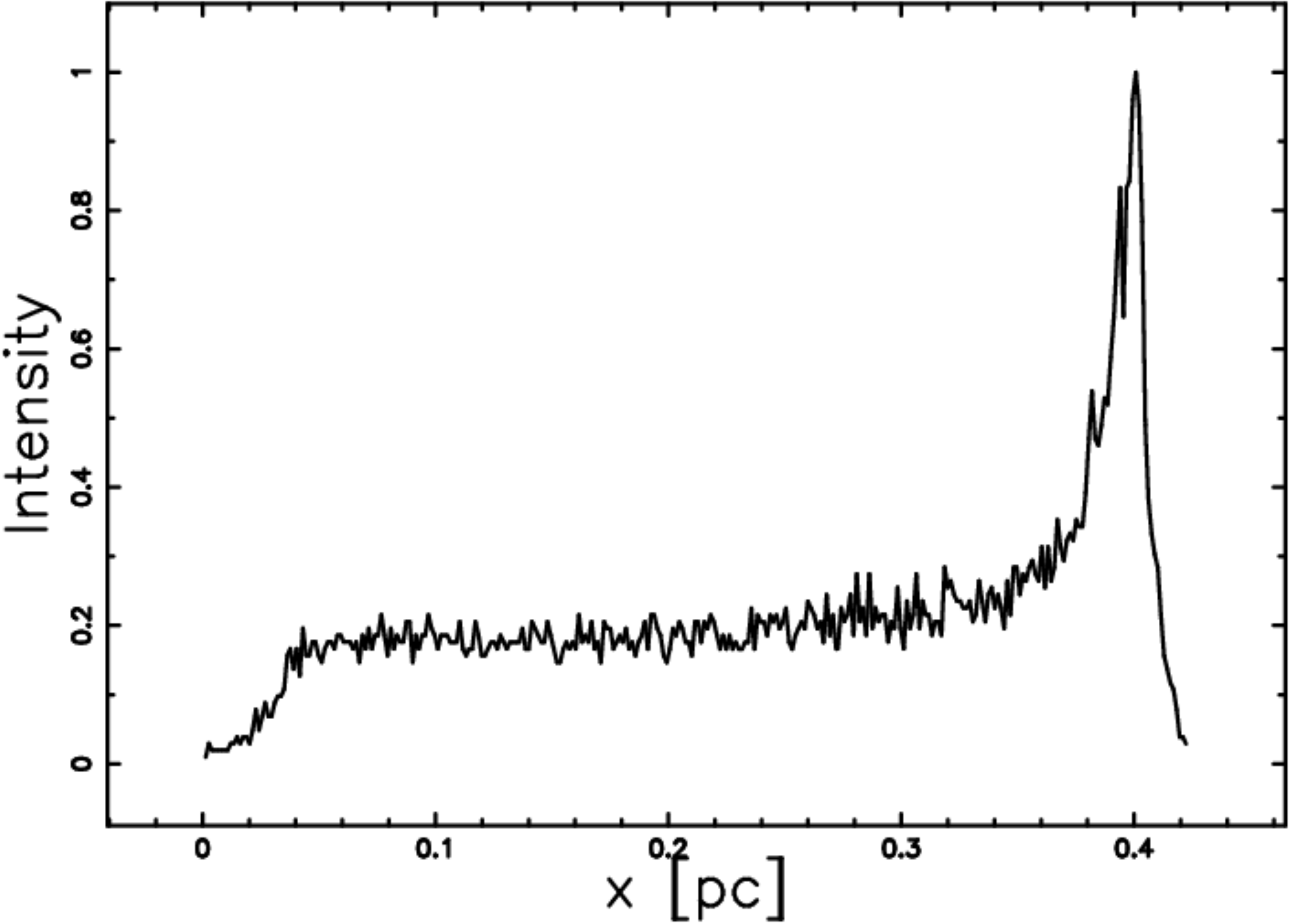}
 \end {center}
\caption {
Behavior of the  intensity or brightness  of HH34 
at the centerline.
Parameters as in Figure~\ref{hh34_complex_zoom} 
}%
 \label{hh34_cut_zoom}
 \end{figure}

The enhancement of the  intensity or brightness  in the previous map
where the jet is bending is due to the particular point of
view of the observer.
Figure ~\ref{hh34_cut_zoom} reports a cut along the centerline
of a jet from which is possible to observe an increase
of a factor $\approx$ 5 in the 
axial  intensity or brightness  otherwise constant
.
An analytical evaluation of such increase is 
reported in Section~\ref{torosec}.

Conversely when the plane of the trajectory 
is perpendicular to the point of view 
of the observer the enhancement in the  intensity or brightness 
of HH34 is not present ,
see Figure~\ref{hh34_complex_proj_2}.

\subsection{Toroidal Model}

\label{torosec}
The curved shape of a jet of finite cross section
is not easy to parametrize.
The torus 
represents a possible model
due to the presence of the small radius 
that characterizes the cross section of the
HH object , $r$ ,
and the great radius $R$ that can be identified 
with the curvature $k$, $k=1/R$ that 
characterizes the 3D trajectory.
The torus 
has the following parametric equations:
\begin{eqnarray}
x = \cos(s) \cdot(R + r \cdot \cos(t)) \nonumber \\
y = \sin(s) \cdot (R + r \cdot \cos(t)) \\
z = r \cdot \sin(t) \quad , \nonumber \\
\nonumber 
\end{eqnarray}
where $ t \in [0,2\pi)$ 
and $ s \in [0,\pi/4)$.

\begin{figure}
 \begin{center}
\includegraphics[width=7cm]{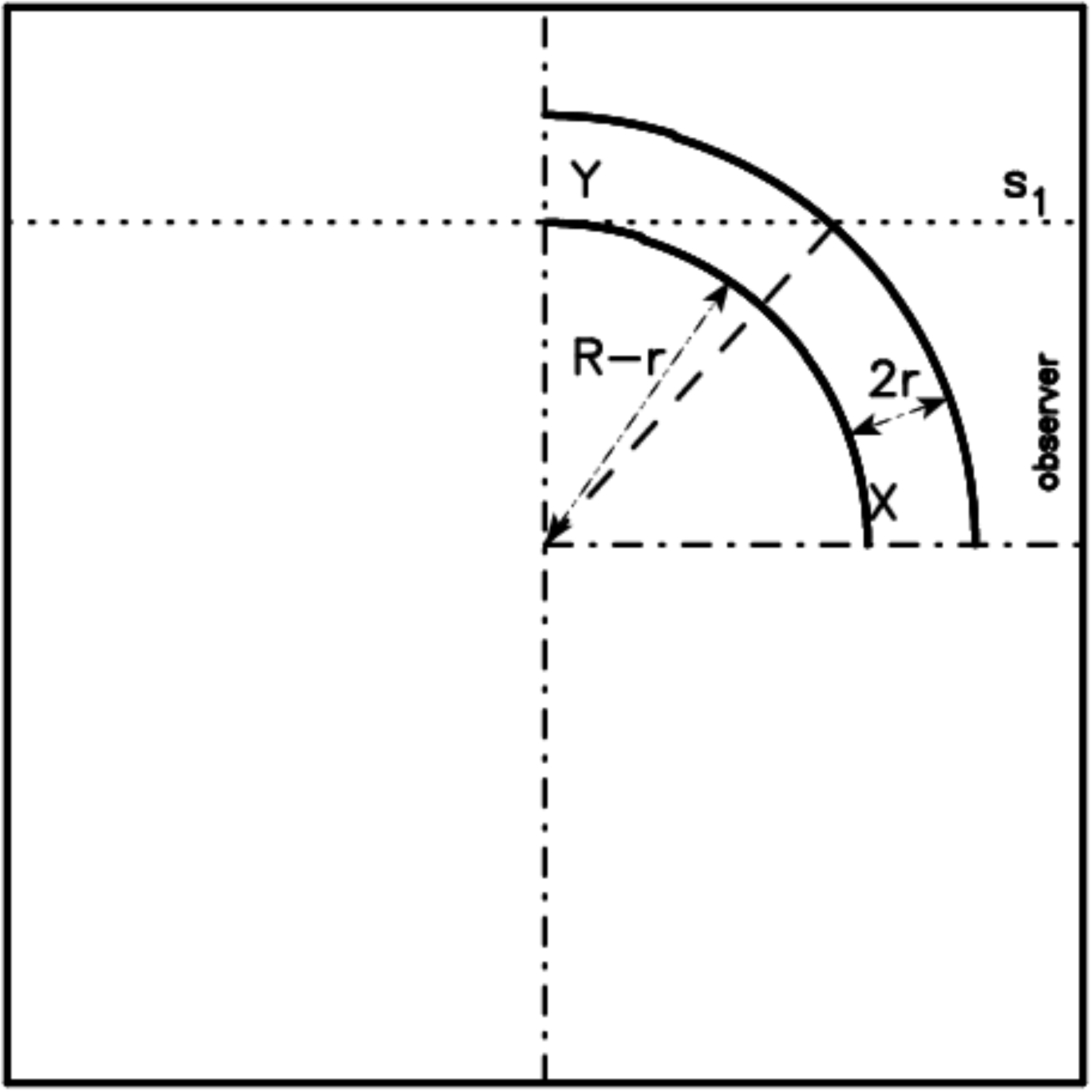}
 \end {center}
\caption {
The section of one fourth of a torus 
is represented through
a circle of radius $R$ and a 
bigger circle of radius $R+2r$.
The observer is situated along the $x$ direction,
the line of sight of maximum length is indicated.
 }%
 \label{torovista}
 \end{figure}

Figure~\ref{torovista} reports a section
in the middle of the torus $z=0$,
from which is possible to see 
that the dotted line presents the 
longest line of sight , $l_{max}$ ,
when the observer is at infinity of the $x-axis$ .
The shortest line of sight is $2r$ .
The maximum enhancement in the presence
of constant number density , $e$ , is
\begin{equation}
e =\frac {l_{max}} {2r}.
\end{equation}
A simple geometrical demonstration gives 
\begin{equation}
e = \sqrt {\frac{R}{r}}
\quad .
\label{fattoree}
\end{equation}
The radius that produces an enhancement $e$ in the
 intensity or brightness 
is therefore
\begin{equation}
\frac{R}{r} = e^2 
\quad .
\end{equation}
As an example an enhancement 
of $e=5$ is produced by a radius 
of curvature 25 times greater in respect to the
HH's radius.

\section{The image from turbulence }

\label{sec_image_turb}
The power released in the turbulent cascade
has the same dimension of the emission coefficient
and therefore
can be considered  the source of  emissivity.
We now consider a linear and a nonlinear relationship
between turbulent power and emission coefficient.

\subsection{Linear correspondence}
 
\label{linear}
It is assumed that the emission coefficient of the HH 
scales as the power released in turbulent kinetic energy,
see equation~(\ref{powerturb}),
\begin{equation}
\epsilon \sim \mathcal{P}
\quad . 
\end{equation}

Due to the additive property  
of the optically thin medium 
along the line of sight, an integral 
operation is performed in order to obtain the 
 intensity or brightness  of emission
\begin{equation}
I(y)=
\int _{0}^{\sqrt {{a}^{2}-{y}^{2}}}
2 \times \epsilon (r) {dz} 
\quad ,
\end{equation}
with $r=\sqrt{z^2 +y^2}$ and $a$ representing the jet radius,
see Figure~\ref{asolo}.

The  intensity or brightness  of emission 
according to formula~(\ref{powerturb}) is 
\begin{eqnarray}
I(y) \sim 
\int _{0}^{\sqrt {{{\it a}}^{2}-{y}^{2}}} Il(z) dz 
\label{iyturb} 
\\
with \quad 
Il(z) =   \nonumber \\
\!4\,{\frac { \left( {{\it 
z}}^{2}+{y}^{2} \right) {A~}^{2}{{\it {b_{\frac{1}{2}}}}}^{8}{{\it a}}^{2}}{
 \left( {{\it {b_{\frac{1}{2}}}}}^{2}+A~{{\it z}}^{2}+A~{y}^{2} \right) ^{6}{x}^{2
} \left( \sqrt {2}-1 \right) \tan \left( \frac{\alpha}{2} \right) }}
\quad .
\nonumber
\end{eqnarray}

\begin{figure}
 \begin{center}
\includegraphics[width=7cm]{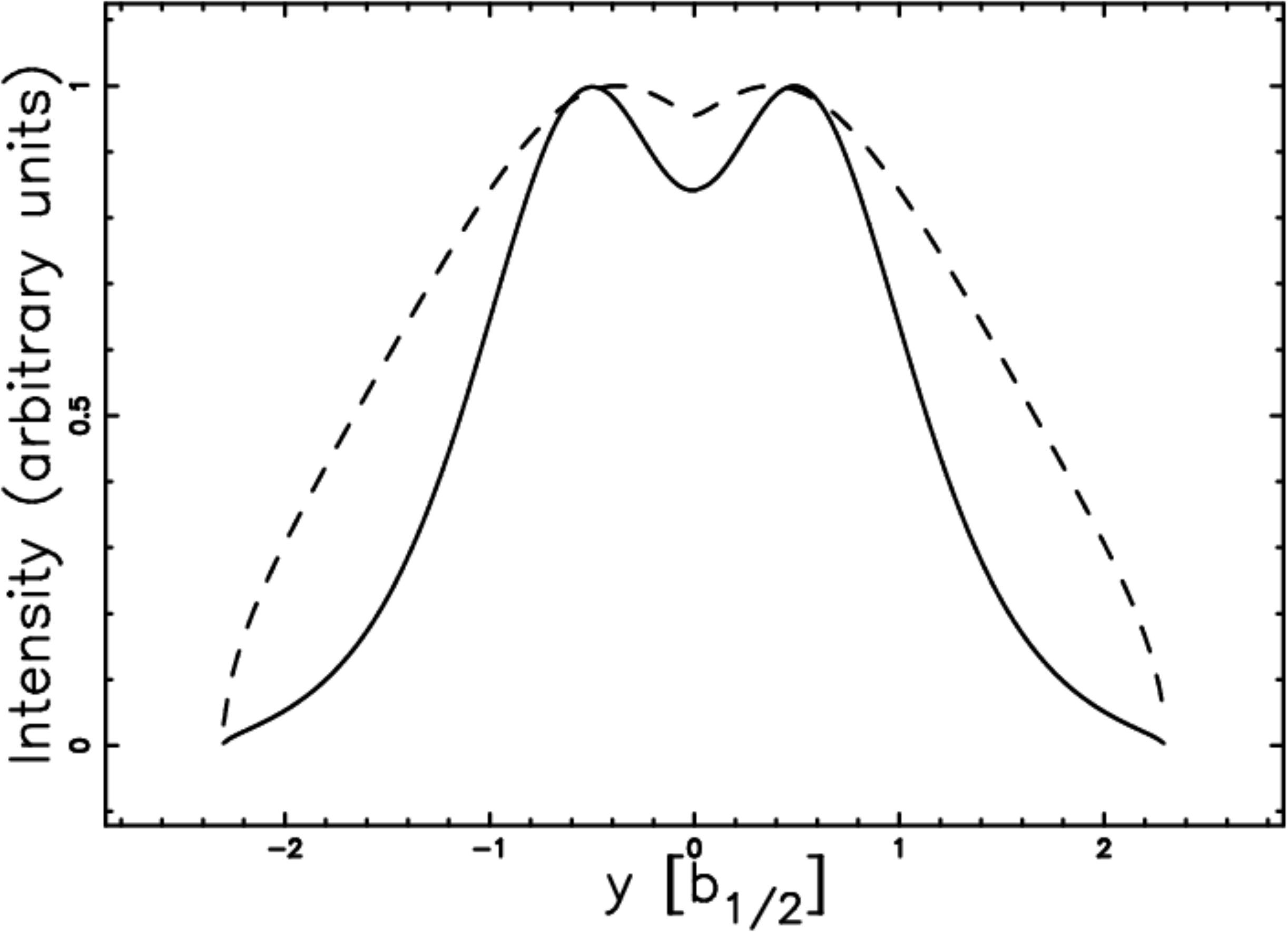}
 \end {center}
\caption {
 Intensity of radiation ${\it I(y)}$ 
 in $b_{\frac{1}{2}}$ units 
across the turbulent jet
 when $z/d=50$ , 
 $k$ =0.54,
 $A~$ = 0.414 
 and 
$\alpha_{deg}=2.0 $ :
full line ( linear correspondence) and dashed line
(non linear correspondence,$f=4$) .
 }%
 \label{integrale_turb}
 \end{figure}

This integral has an analytical solution but it is
complicated and 
therefore Figure~\ref{integrale_turb} 
only shows 
the numerical integration
which presents  a characteristic shape 
on the top of the blob
called the "valley on the top" .
The maximum of this integral is at the point 
$y \approx 0.49{b_{\frac{1}{2}}}$ and the 
value of  intensity or brightness  at the maximum is $1.18$ times
the value at the point $y=0$ ( the center of the jet).
The near infrared images of HH~110 jet
were interpreted as due to low velocity shocks produced by turbulent
processes,
see 
 \cite{Noriega-Crespo1996}.
The spatial  intensity or brightness  distribution of $H_2$ , 
$H_{\alpha}$ and
$[S_{II}]_{6717/31}$ perpendicular to the flow axis and along 
the cross section of knots in $HH110$ has a behavior that 
can be approximated by a Gaussian distribution , 
see Figure~4 in \cite{Noriega-Crespo1996}.
In one case , $H_{\alpha}$ in knot P of HH110
in Figure~5 in \cite{Noriega-Crespo1996}
, it 
is possible to see a bump near the maximum of the  intensity or brightness 
in the transversal direction. 
A second  observation that presents 
a bimodal profile 
is
the spatial $H_{\alpha}$ intensity distribution
through the cross section of knot $I+J$ 
of HH110 visible in Figure 10  in \cite{Riera2003}.
A third observations is the $[SII]+[NII]$ profile
in blob 1 of HH110  as in  Figure 17 of \cite{Hartigan2009}.

These three cases  can be considered an observational evidence 
of the physical effect previously named " valley on the top" .

\begin{figure}
 \begin{center}
\includegraphics[width=7cm]{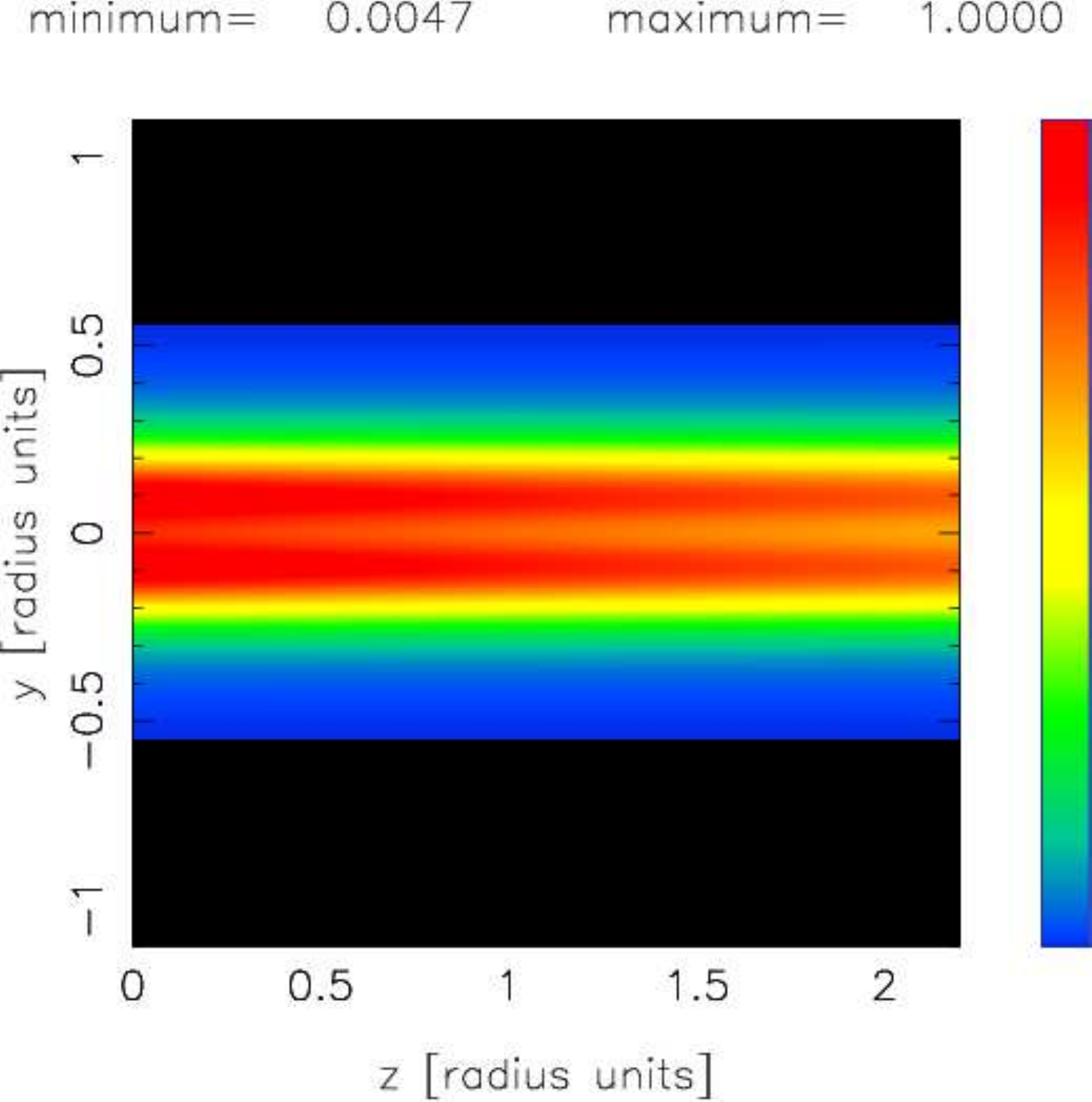}
 \end {center}
\caption {
Theoretical 2D map of the  surface
  brightness  of emission
corresponding to a $2.2~arcsec$ of HH1
around $z/d=110$. 
The parameters are 
$k$ =0.54,
$A$ = 0.414 
and 
$\alpha_{deg}=2.08$ .
The integral operation is performed on a cubic grid 
of $400^3$ pixels.
 }%
 \label{hh1_zoom}
 \end{figure}

Is also possible to build a 2D map 
of the  surface  brightness  
of emission computed according to the integral of 
equation(\ref{iyturb}) in a conical jet as
HH1 and Figure~\ref{hh1_zoom} reports 
such a map.

\begin{figure}
 \begin{center}
\includegraphics[width=7cm]{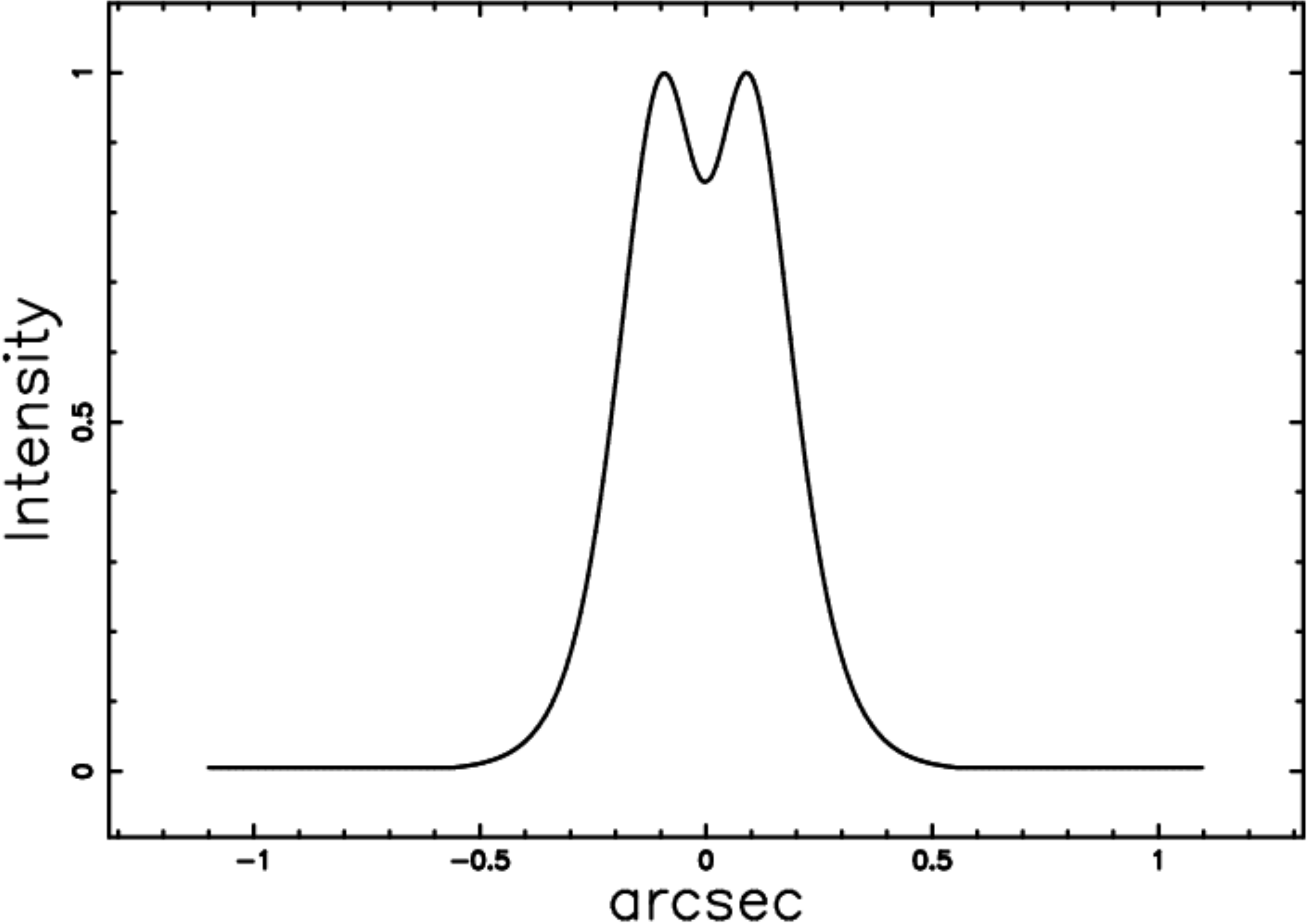}
 \end {center}
\caption {
Intensity of HH1 represented
through a cut in the y-direction , parameters as 
in Figure~\ref{hh1_zoom}. 
 }%
 \label{hh1_transverse}
 \end{figure}

In this case the values of emissivity are memorized on 
a 3D grid made by $(pixels)^3$ points.
The integral is represented by a sum along the line of
sight and Figure~\ref{hh1_transverse} reports 
a cut in the middle.

\subsection{Non linear correspondence}

From a careful inspection of formula~(\ref{powerturb})
we see that the local 
power released in the turbulent cascade at $r=b_{\frac{1}{2}}$
scale as $\sim (\frac{1}{x})^4$ .
In order to have a constant  intensity or brightness  along the jet
we now consider the case
\begin{equation}
\epsilon \sim ({\mathcal{P}})^{1/f}
\quad , 
\end{equation}
which means that the  intensity or brightness  
scales as 
\begin{equation}
I(x) \sim d(x) (\frac{1}{x^{4}})^{1/f} 
\sim (\frac{1}{x}) ^{\frac{4}{f} -1}
\quad .
\end{equation}

The  intensity or brightness  of emission 
is 
\begin{eqnarray}
I(y) \sim 
\int _{0}^{\sqrt {{{\it a}}^{2}-{y}^{2}}}
 In(z) dz  
\label{iyturbf} 
\\ 
with \quad In(z)=   \nonumber \\
(\!4\,{\frac { \left( {{\it 
z}}^{2}+{y}^{2} \right) {A~}^{2}{{\it {b_{\frac{1}{2}}}}}^{8}{{\it a}}^{2}}{
 \left( {{\it {b_{\frac{1}{2}}}}}^{2}+A~{{\it z}}^{2}+A~{y}^{2} \right) ^{6}{x}^{2
} \left( \sqrt {2}-1 \right) \tan \left( \frac{\alpha}{2} \right) }}
)^{\frac{1}{f}} 
 .
\nonumber
\end{eqnarray}
The numerical result for $f=4$ , constant  intensity or brightness  
along the main direction $x$, 
is reported in 
Figure~\ref{integrale_turb} as a dashed line.

\section{Conclusions}

{\bf Law of motion}
The two theories here considered are based on the 
behavior  of the centerline velocity 
, see the astrophysical equations~(\ref{u_astro}) 
and (\ref{u_astro_simple}) .
From the two previous equations is 
possible to deduce the law of motion 
in presence of a stationary state,
see the astrophysical equations~(\ref{x_astro}) 
and (\ref{x_astro_simple}) .
On the way the flow rate of mass and the 
flow rate of energy (the mechanical luminosity)
 are also derived 
, see the astrophysical equations~(\ref{mass_astro}) 
,(\ref{mass_astro_simple}) 
,(\ref{luminosity_astro}) ,
and (\ref{luminosity_astro_simple}).

{\bf Images}
The analysis  of the  intensity or brightness  of a HH object 
has been split in three theoretical parts
corresponding to three observable cases.
\begin{enumerate}
\item 
{\it Transversal cut} 
We have analyzed the case 
of constant number density , see 
equation~(\ref{icylindera}), and the 
case of emissivity connected with the power
released in the turbulent cascade,
see equations (\ref{iyturb}) and  (\ref{iyturbf}).
The  intensity or brightness  from turbulent cascade originates
a curious effect at the center of the jet
named  "valley on the top" .
\item 
{\it Longitudinal  cut}
Through a parametrization of the number density 
is possible to fit the theoretical and the observed
 intensity or brightness  ,  see equation (\ref{ilongitudinal}) .
\item 
{\it 2D map }
The details of the HH's image can be simulated
imposing an arbitrary  point of view of the observer.
The enhancement in   intensity or brightness   is explored 
from a numerical point of view ,
see Figure  (\ref{hh34_complex_zoom}) and 
 (\ref{hh34_cut_zoom}).
An analytical  explanation 
of the enhancement in   intensity or brightness  is derived 
from the geometrical  properties of the
torus ,
see formula~(\ref{fattoree}).
Is interesting to underline that the "torus effect"
replaces the concept of bow shock.
 that is 
often used in order to explain the intensity
enhancement along the HHs, see
\cite{Krist1999,Smith2003}.
\end{enumerate}


\end{document}